\documentclass[draftclsnofoot,onecolumn]{IEEEtran}
\pdfoutput=1

\usepackage{graphicx}
\usepackage[cmex10]{amsmath} 
\usepackage{amssymb}
\usepackage[tight,footnotesize]{subfigure}
\usepackage{bm}
\usepackage{cite}

\renewcommand{\Re}{\mathbb{R}}

\newtheorem{definition}{Definition}

\newtheorem{theorem}{Theorem}

\newcommand{\beq}{\begin{equation}}
\newcommand{\eeq}{\end{equation}}
\newcommand{\stexp}{\mbox{$\mathbb{E}$}}    

\newcommand{\ones}{\ensuremath{\vec{1}}}

\newcommand{\Prob}{\ensuremath{\mathbb{P}}}

\newcommand{\xave}{\ensuremath{\bar{x}_{\operatorname{ave}}}}

\usepackage{paralist}

\begin{document}

\def\titletext{Gossip Algorithms for Distributed Signal Processing}
\title{\titletext}

\author{Alexandros~G.~Dimakis, 
        Soummya~Kar, 
        	Jos\'{e}~M.F.~Moura, 
	Michael~G.~Rabbat, 
        and~Anna~Scaglione, 
\thanks{Manuscript received November 16, 2009; revised March 26, 2010.} 
\thanks{A.G. Dimakis is with the Department
of Electrical and Computer Engineering, University of Southern California, Los Angeles,
CA, 90089 USA \{e-mail: dimakis@usc.edu.\}}%
\thanks{S. Kar and J.M.F. Moura are with the Department
of Electrical and Computer Engineering, Carnegie Mellon University, Pittsburgh,
PA, 15213 USA \{e-mail: soummyak@andrew.cmu.edu; moura@ece.cmu.edu.\}}%
\thanks{M.G. Rabbat is with the Department
of Electrical and Computer Engineering, McGill University, Montr\'{e}al,
QC, H3A 2A7 CANADA \{e-mail: michael.rabbat@mcgill.ca.\}}%
\thanks{A. Scaglione is with the Department
of Electrical and Computer Engineering, University of California, Davis,
CA, 95616 USA \{e-mail: ascaglione@ucdavis.edu.\}}%
\thanks{The work of Kar and Moura was partially supported by the NSF under grants 
ECS-0225449 and CNS-0428404, and by the Office of Naval Research under 
MURI N000140710747.  The work of Rabbat was partially supported by the NSERC under grant RGPIN 341596-2007, by MITACS, and by FQRNT under grant 2009-NC-126057.  The work of Scaglione is supported by the NSF under grant CCF-0729074.}%
}

\def\headertext{Dimakis \emph{et al.}: \titletext}
\markboth{\headertext}{\headertext}%

\maketitle

\begin{abstract}
Gossip algorithms are attractive for in-network processing in sensor networks because they do not require any specialized routing, there is no bottleneck or single point of failure, and they are robust to unreliable wireless network conditions.  Recently, there has been a surge of activity in the computer science, control, signal processing, and information theory communities, developing faster and more robust gossip algorithms and deriving theoretical performance guarantees.  This article presents an overview of recent work in the area.  We describe convergence rate results, which are related to the number of transmitted messages and thus the amount of energy consumed in the network for gossiping.  We discuss issues related to gossiping over wireless links, including the effects of quantization and noise, and we illustrate the use of gossip algorithms for canonical signal processing tasks including distributed estimation, source localization, and compression.
\end{abstract}

\section{Introduction}
\label{sec:intro}

Collaborative in-network processing is a major tenet of wireless sensor networking, and has received much attention from the signal processing, control, and information theory communities during the past decade \cite{zhao04}.  Early research in this area considered applications such as detection, classification, tracking, and pursuit \cite{Zhao03,Sinopoli03,Chong03,Brooks03}.  By exploiting local computation resources at each node, it is possible to reduce the amount of data that needs to be transmitted out of the network, thereby saving bandwidth and energy, extending the network lifetime, and reducing latency.

In addition to having on-board sensing and processing capabilities, the archetypal sensor network node is battery-powered and uses a wireless radio to communicate with the rest of the network.  Since each wireless transmission consumes bandwidth and, on common platforms, also consumes considerably more energy than processing data locally \cite{pottie00, shnayder04}, reducing the amount of data transmitted can significantly prolong battery life.  In applications where the phenomenon being sensed varies slowly in space, the measurements at nearby sensors will be highly correlated.  In-network processing can compress the data to avoid wasting transmissions on redundant information.  In other applications, rather than collecting data from each node, the goal of the system may be to compute a function of the data such as estimating parameters, fitting a model, or detecting an event.  In-network processing can be used to carry out the computation within the network so that, instead of transmitting raw data to a fusion center, only the results of the computation are transmitted to the end-user.  In many situations, in-network computation leads to considerable energy savings over the centralized approach \cite{yu04,rabbat04}.

Many previous approaches to in-network processing assume that the network can provide specialized routing services.  For example, some schemes require the existence of a cyclic route through the network that passes through every node precisely one time\footnote{This is a Hamiltonian cycle, in graph-theoretic terms.} \cite{rabbat04,blatt06,son05}.  Others are based on forming a spanning tree rooted at the fusion center or information sink, and then aggregating data up the tree \cite{yu04,Ciancio06,Ratnasamy03}.  Although using a fixed routing scheme is intuitive, there are many drawbacks to this approach in wireless networking scenarios.  Aggregating data towards a fusion center at the root of a tree can cause a bottleneck in communications near the root and creates a single point of failure.  Moreover, wireless links are unreliable, and in dynamic environments, a significant amount of undesirable overhead traffic may be generated just to establish and maintain routes.

\subsection{Gossip Algorithms for In-Network Processing}

This article presents an overview of gossip algorithms and issues related to their use for in-network processing in wireless sensor networks.  Gossip algorithms have been widely studied in the computer science community for information dissemination and search \cite{karp00,kempe03,levis04}.  More recently, they have been developed and studied for information processing in sensor networks.  They have the attractive property that no specialized routing is required.  Each node begins with a subset of the data in the network.  At each iteration, information is exchanged between a subset of nodes, and then this information is processed by the receiving nodes to compute a local update.

Gossip algorithms for in-network processing have primarily been studied as solutions to \emph{consensus problems}, which capture the situation where a network of agents must achieve a consistent opinion through local information exchanges with their neighbors.  Early work includes that of Tsitsiklis et al.~\cite{Tsitsiklis84,Tsitsiklis86}.  Consensus problems have arisen in numerous applications including: load balancing~\cite{cybenko89}; alignment, flocking, and multi-agent collaboration~\cite{jadbabaielinmorse03,SensNets:Olfati04}; vehicle formation~\cite{faxmurray04}, tracking and data fusion~\cite{salig:06}, and distributed inference~\cite{tsp06-K-A-M}.

The canonical example of a gossip algorithm for information aggregation is a randomized protocol for distributed averaging.  The problem setup is such that each node in a $n$-node network initially has a scalar measurement value, and the goal is to have every node compute the average of all $n$ initial values -- often referred to as the average consensus.  In pairwise randomized gossiping~\cite{boyd06}, each node maintains an estimate of the network average, which it initializes with its own measurement value.  Let $x(t)$ denote the vector of estimates of the global averages after the $t^{th}$ gossip round, where $x(0)$ is the vector of initial measurements; that is, $x_i(t)$ is the estimate\footnote{Throughout, we will sometimes alternatively refer to the estimates $x_i(t)$ as states, and to nodes as agents.} at node $i$ after $t$ iterations.  In one iteration, a randomly selected pair of neighboring nodes in the network exchange their current estimates, and then update their estimates by setting $x_i(t+1) = x_j(t+1) = \big(x_i(t) + x_j(t)\big)/2$.  A straightforward analysis of such an algorithm shows that the estimate at each node are guaranteed to converge to the average, $x_{ave} = \frac{1}{n}\sum_{i=1}^n x_i(0)$, as long as the network is connected (information can flow between all pairs of nodes), and as long as each pair of neighboring nodes gossips frequently enough; this is made more precise in Section~\ref{sec:rates} below.  Note that the primitive described above can be used to compute any function of the form $\sum_{i=1}^n f_i\big(x_i(0)\big)$ by properly setting the initial value at each node, and while this is not the most general type of query, many useful computations can be reduced in this form as will further be highlighted in Sections~\ref{sec:apps}~and~\ref{sec:conc}.

Gossip algorithms can be classified as being randomized or deterministic.  The scheme described above is randomized and asynchronous, since at each iteration a random pair of nodes is active.  In deterministic, synchronous gossip algorithms, at each iteration node $i$ updates $x_i(t+1)$ with a convex combination of its own values and the values received from all of its neighbors, e.g., as discussed in \cite{Xiao03}.  Asynchronous gossip is much better suited to wireless sensor network applications, where synchronization itself is a challenging task.  Asynchronous gossip can be implemented using the framework described in \cite{Tsitsiklis86,bertsekasPdc}.  Each node runs an independent Poisson clock, and when node $i$'s clock ``ticks", it randomly selects and gossips with one neighbor.  In this formulation, denoting the probability that node $i$ chooses a neighbor $j$ by $P_{i,j}$, conditions for convergence can be expressed directly as properties of these probabilities.  Gossip and consensus algorithms have also been the subject of study within the systems and control community, with a focus on characterizing conditions for convergence and stability of synchronous gossiping, as well as optimization of the algorithm parameters $P_{i,j}$; see the excellent surveys by Olfati-Saber and Murray~\cite{olfati07}, and Ren et al.~\cite{ren07}, and references therein.

\subsection{Paper Outline}

Our overview of gossip algorithms begins on the theoretical side and progresses towards sensor network applications.  Each gossip iteration requires wireless transmission and thus consumes valuable bandwidth and energy resources.  Section~\ref{sec:rates} discusses techniques for bounding rates of convergence for gossip, and thus the number of transmissions required.  Because standard pairwise gossip converges slowly on wireless network topologies, a large body of work has focused on developing faster gossip algorithms for wireless networks, and this work is also described.  When transmitting over a wireless channel, one must also consider issues such as noise and coding.  Section~\ref{sec:wireless} discusses the effects of finite transmission rates and quantization on convergence of gossip algorithms.  Finally, Section~\ref{sec:apps} illustrates how gossip algorithms can be applied to accomplish distributed signal processing tasks such as distributed estimation and compression.

\section{Rates of Convergence and Faster Gossip}
\label{sec:rates}

Gossip algorithms are iterative, and the number of wireless messages transmitted is proportional to the number of iterations executed.  Thus, it is important to characterize rate of convergence of gossip and to understand what factors influence these rates.  This section surveys convergence results, describing the connection between the rate of convergence and the underlying network topology, and then describes developments that have been made in the area of fast gossip algorithms for wireless sensor networks.

\subsection{Analysis of Gossip Algorithms}
In pairwise gossip, only two nodes exchange information at each iteration.  More generally, a subset of nodes may average their information.  All the gossip algorithms that we will be interested in can be described by an equation of the form
\beq
x(t+1)= W(t) x(t), \label{distrEq}
\eeq
where $W(t)$ are randomly selected averaging matrices, selected independently across time, and $x(t) \in \Re^n$ is the vector of gossip states after $t$ iterations.
When restricted to pairwise averaging algorithms, in each gossip round only the values of two nodes $i,j$
are averaged (as in~\cite{boyd06}) and the corresponding $W(t)$ matrices have
$1/2$ in the coordinates $(i,i), (i,j),(j,i),(j,j)$ and a diagonal identity for every other
node. When pairwise gossip is performed on a graph $G = (V,E)$, only the matrices that average nodes that are neighbors on $G$ (i.e., $i,j \in E$) are selected with non-zero probability. More generally, we will be interested in matrices that average sets of node values
and leave the remaining nodes unchanged. A matrix $W(t)$ acting on a vector $x(t)$ is \emph{set averaging matrix} for a set $S$ of nodes, if
\beq
x_i(t+1)=  \frac{1}{|S|} \sum_{i\in S} x(t) \quad , i\in S,
\eeq  
and $x_i(t+1)= x_i(t)$, $i \notin S$.  Such matrices therefore have entry $1/|S|$ at the coordinates corresponding to the set $S$ and
a diagonal identity for all other entries.

It is therefore easy to see that all such matrices will have the following properties:
\beq
\label{conditions}
\begin{cases}
 \ones^{T}W(t)=\ones^{T}\\
 W(t)\ones=\ones,
\end{cases}
\eeq
which respectively ensure
that the average is preserved at every iteration, and that $\ones$, the vector of ones, is a fixed point.
Further, any set averaging matrix $W$ is symmetric and doubly stochastic.  A matrix is doubly stochastic if its rows sum to unity, and its columns also sum to unity, as implied in \eqref{conditions}.  The well-known Birkhoff--von Neumann Theorem states that a matrix is doubly stochastic if and only if it is a convex combination of permutation matrices.  In the context of gossip, the only permutation matrices which contribute to the convex combination are those which permute nodes in $S$ to other nodes in $S$, and keep all other nodes not in $S$ fixed.
The matrix $W$ must also be a projection matrix; i.e.,
$W^2= W$ since averaging the same set twice no longer changes the vector $x(t)$.
It then follows that $W$ must also be positive-semidefinite.

We are now ready to understand the evolution of the estimate vector $x(t)$ through the
product of these randomly selected set averaging matrices:
\beq
\label{evolution1}
x(t+1)= W(t) x(t) = \prod_{k=0}^t W(k) \,\, x(0).
\eeq
Since $W(t)$ are selected independently across time, $\stexp W(t)= \stexp W(0)$ and
we can drop the time index and simply refer to the expected averaging matrix $\stexp W$,
which is the average of symmetric, doubly stochastic, positive semidefinite matrices and therefore also has these properties.
The desired behavior is that $x(t+1) \rightarrow x_{ave}\ones$ that is equivalent to asking that
\beq
\prod_{k=0}^t W(k) \rightarrow \frac{1}{n} \ones \ones^T.
\eeq
\subsection{Expected behavior}
We start by looking at the expected evolution of the random vector $x(t)$ by taking
expectations on both sides of (\ref{evolution1}):
\beq
\stexp x(t+1) = \stexp \left( \prod_{k=0}^t W(k) \right)  \,\, x(0)= ( \stexp W )^{t+1} x(0),
\eeq
where the second equality is true because the matrices are selected independently.
Since $\stexp W$ is a convex combination of the matrices $W(t)$ which all satisfy the conditions \eqref{conditions}, it is clear that $\stexp W$ is also a doubly stochastic matrix.  We can see that the expected evolution of
the estimation vector follows a Markov chain that has the $\xave \ones$ vector as its stationary distribution.  In other words, $\ones$ is an eigenvector of $\stexp W$ with eigenvalue $1$.
Therefore if the Markov chain corresponding to $\stexp W$ is irreducible and aperiodic, our estimate vector will converge in expectation to the desired average.
Let $\lambda_2(\stexp[W])$ be the second largest eigenvalue of $\stexp W$. If condition (\ref{conditions}) holds and if $\lambda_2(\stexp[W])<1$, then $x(t)$ converges to $x_{ave}\ones$ in expectation and in mean square. Further precise conditions for convergence in expectation and in mean square can be found in \cite{BoydGPS:04cdc}.

\subsection{Convergence rate}
The problem with the expectation analysis is that it gives no estimate on the rate of convergence,
a key parameter for applications. Since the algorithms are randomized, we need
to specify what we mean by convergence. One notion that yields clean
theoretical results involves defining convergence as the first time where the normalized error
is small with high probability, and controlling both error and probability with one parameter, $\epsilon$.
\begin{definition}\emph{ $\epsilon$-averaging time $T_{ave}(\epsilon)$}.
Given $\epsilon > 0$, the $\epsilon$-averaging time is the earliest
gossip round in which the vector $x(k)$ is $\epsilon$ close to the normalized
true average with probability greater than $1 - \epsilon$:
\begin{equation}
\label{ave_time_definition} T_{ave}(\epsilon)=\sup_{x(0)} \inf_{t = 0,1,2 \ldots}
\left\{\Prob \left( \frac{ \|x(t)- x_{ave} \ones \|}{\|x(0)\|} \geq
\epsilon \right) \leq \epsilon \right\}.
\end{equation}
\end{definition}
Observe that the convergence time is defined for the worst case over the initial vector of measurements $x(0)$. This definition was first used in~\cite{boyd06} (see also~\cite{FagnaniJS}
for a related analysis).

The key technical theorem used in the analysis of gossip algorithms is the following connection
between the averaging time and the second largest eigenvalue of $\stexp W$:
\begin{theorem}
For any gossip algorithm that uses set-averaging matrices and converges in expectation,
the averaging time is bounded by
\beq
 \label{Boyd_bound} T_{ave}(\epsilon, \stexp W)\leq
\frac{3 \log \epsilon^{-1}}{\log\left(\frac{1}{\lambda_2 (\stexp W)}\right)} \leq
\frac{3 \log \epsilon^{-1}}{1-\lambda_2(\stexp W)}. \eeq
\end{theorem}
This theorem is a slight generalization of Theorem 3 from~\cite{boyd06} for non-pairwise averaging
gossip algorithms. There is also a lower bound of the same order, which implies that\footnote{Because our primary interest is in understanding scaling laws---how many messages are needed as the network size grows---our discussion centers on the order-wise behavior of gossip algorithms.  Recall the Landau or ``big O'' notation: a function $f$ is asymptotically bounded above by $g$, written $f(n) = O(g(n))$, if there exist constants $N > 0$ and $c_1 > 0$ such that $f(n) \le c_1 \, g(n)$ for all $n \ge N$;  $f$ is asymptotically bounded below by $g$, written $f(n) = \Omega(g(n))$, if there exist constants $c_2 > 0$ and $N> 0$ such that $f(n) \ge c_2 \, g(n)$ for $n \ge N$; and $f$ is asymptotically bounded above and below by $g$, written $f(n) = \Theta(g(n))$, if $c_2\, g(n) \le f(n) \le c_1 \, g(n)$ for all $n \ge N$.} $T_{ave}(\epsilon, \stexp W)=\Theta(\log{\epsilon^{-1}}/(1-\lambda_{2}(\stexp W)))$.

The topology of the network influences the convergence time
of the gossip algorithm, and using this theorem this is precisely quantified; the matrix $\stexp[W]$ is completely specified by the network topology and the selection
probabilities of which nodes gossip.
The rate at which the \emph{spectral gap} $1-\lambda_2(\stexp[W])$ approaches zero, as $n$ increases, controls the $\epsilon$-averaging time $T_{ave}$.
The spectral gap is related to the mixing time (see, e.g.,~\cite{Sinclair}) of a random walk on the network topology. Roughly, the gossip averaging time is the mixing time of the simple random walk on the graph times a factor
of $n$.
One therefore would like to understand how the spectral gap scales for different models of networks and gossip algorithms.

This was first analyzed for the complete graph and uniform pairwise
gossiping~\cite{kempe03,BoydGPS:04cdc,boyd06}.
For this case it was shown that $\lambda_2(\stexp[W])=1-1/n$ and therefore, $T_{ave}=\Theta(n\log\epsilon^{-1})$. Since only nearest neighbors interact, each gossip round costs two transmitted messages, and therefore,
$\Theta (n \log \epsilon^{-1})$ gossip messages need to be exchanged to converge to the global average within $\epsilon$ accuracy.
This yields $\Theta(n \log n)$ messages to have a vanishing error with probability $1/n$,
an excellent performance for a randomized algorithm with no coordination that averages $n$ nodes on the complete graph.
For other well connected graphs (including expanders and small world graphs),
uniform pairwise gossip converges very quickly, asymptotically requiring the same number of messages
($\Theta (n \log \epsilon^{-1})$) as the complete graph. Note that any algorithm that averages $n$ numbers with a constant error and constant probability of success should require $\Omega(n)$ messages.

If the network topology is fixed, one can ask what is the selection of pairwise gossiping probabilities that maximizes the convergence rate (i.e. maximizes the spectral gap).
This problem is equivalent to designing a Markov chain which approaches stationarity optimally fast and, interestingly, it can be formulated as a semidefinite program (SDP) which can be solved efficiently~\cite{Boyd03fastestmixing,Xiao03,boyd06}.
Unfortunately, for random geometric graphs\footnote{The family of random geometric graphs with $n$ nodes and connectivity radius $r$, denoted $\mathcal{G}(n,r)$, is obtained by placing $n$ nodes uniformly at random in the unit square, and placing an edge between two nodes if their Euclidean distance is no more than $r$.  In order to process data in the entire network, it is important that the network be \emph{connected} (i.e., there is a route between every pair of nodes).  A fundamental result due to Gupta and Kumar~\cite{gupta98} states that the critical connectivity threshold for $\mathcal{G}(n,r)$ is $r_{con}(n) = \Theta(\sqrt{\frac{\log n}{n}})$; that is, if $r$ does not scale as fast as $r_{con}(n)$, then the network is not connected with high probability, and if $r$ scales at least as fast as $r_{con}(n)$, then the network is connected with high probability.  Throughout this paper, when using random geometric graphs it is implied that we are using $\mathcal{G}(n, r_{con}(n))$, in order to ensure that information flows across the entire network.} and grids, which are the relevant topologies for large wireless ad-hoc and sensor networks, even the optimized version of pairwise gossip
is extremely wasteful in terms of communication requirements.
For example for a grid topology, the number of required messages scales like $\Theta(n^2 \log \epsilon^{-1})$~\cite{boyd06,DimakisSW:06ipsn}. Observe that this is of the same
order as the energy required for every node to flood its estimate to all other nodes.
On the contrary, the obvious solution of averaging numbers on a spanning tree
and flooding back the average to all the nodes requires only $O(n)$ messages.
Constructing and maintaining
a spanning tree in dynamic and ad-hoc networks introduces significant overhead and complexity, but a quadratic number of messages is a high price to pay for fault tolerance.

\subsection{Faster Gossip Algorithms}

Pairwise gossip converges very slowly on grids and random geometric graphs because of
its diffusive nature. Information from nodes is essentially performing random walks, and,
as is well known, a random walk on the two-dimensional lattice has to perform $d^2$ steps to cover distance $d$. One approach to gossiping faster is to modify the algorithm so that there is some directionality in the
underlying diffusion of information. Assuming that nodes have knowledge of their geographic
location, we can use a modified algorithm called \emph{geographic gossip}~\cite{DimakisSW:06ipsn}. The idea of geographic gossip is to combine gossip with greedy geographic routing towards a randomly selected location.
If each node has knowledge of its own location and under some mild
assumptions on the network topology, greedy geographic routing can be used to build
an \emph{overlay network} where \emph{any pair} of nodes can communicate. The overlay network is a complete graph on which pairwise uniform gossip converges with $\Theta (n \log \epsilon^{-1})$ iterations. At each iteration, we perform greedy routing, which costs $\Theta(\sqrt{n/\log n})$ messages on a random geometric graph (also the order of the diameter of the network). In total, geographic gossip thus requires
$\Theta(n^{1.5}\log \epsilon^{-1}/\sqrt{\log n})$ messages. The technical part of the analysis
involves understanding how this can be done with only local information: assuming that
each node only knows their own location, routing towards a randomly selected location
is not identical to routing towards a randomly selected node. If the nodes are evenly spaced,
however, these two processes are almost the same and the $\Theta(n^{1.5})$ message scaling
still holds~\cite{DimakisSW:06ipsn}.

Li and Dai~\cite{LiDaiIT,li09} recently proposed \emph{Location-Aided Distributed Averaging (LADA)}, a scheme that uses partial locations and Markov chain lifting to create fast gossiping algorithms. Lifting of gossip algorithms is based on the seminal work of Diaconis et al.~\cite{diaconis2000analysis} and Chen et al.~\cite{chen1999lifting} on lifting Markov chain samplers to accelerate convergence rates.  The basic idea is to lift the original chain to one with additional states; in the context of gossiping, this corresponds to replicating each node and associating all replicas of a node with the original.  LADA creates one replica of a node for each neighbor and associates the policy of a node given it receives a message from the neighbor with that particular lifted state.  In this manner, LADA suppresses the diffusive nature of reversible Markov chains that causes pairwise randomized gossip to be slow.  The cluster-based LADA algorithm performs slightly better than geographic gossip, requiring $\Theta(n^{1.5}\log \epsilon^{-1}/{(\log n)}^{1.5})$ messages for random geometric graphs.
While the theoretical machinery is different, LADA algorithms also use directionality to accelerate gossip, but can operate even with partial location information and have smaller total delay compared to geographic gossip, at the cost of a somewhat more complicated algorithm.  A related scheme based on lifting was proposed concurrently by Jung, Shah, and Shin~\cite{jung07}.
Mosk-Aoyama and Shah~\cite{MoskAoyamaS:05gossip} use an algorithm based on the work of Flajolet and Martin \cite{FlajoletM:85database} to compute averages and bound the averaging time in terms of a ``spreading time'' associated with the communication graph, with a similar scaling
for the number of messages on grids and RGGs.

Just as algorithms based on lifting incorporate additional memory at each node (by way of additional states in the lifted Markov chain), another collection of algorithms seek to accelerate gossip computations by having nodes remember a few previous state values and incorporate these values into the updates at each iteration.  These memory-based schemes can be viewed as predicting the trajectory as seen by each node, and using this prediction to accelerate convergence.  The schemes are closely related to shift-register methods studied in numerical analysis to accelerate linear system solvers.  The challenge of this approach is to design local predictors that provide speedups without creating instabilities.  Empirical evidence that such schemes can accelerate convergence rates is shown in \cite{Cao06}, and numerical methods for designing linear prediction filters are presented in \cite{Kok09,johansson08}.  Recent work of Oreshkin et al.~\cite{Oreshkin09} shows that improvements in  convergence rate on par with of geographic gossip are achieved by a deterministic, synchronous gossip algorithm using only one extra tap of memory at each node.  Extending these theoretical results to asynchronous gossip algorithms remains an open area of research.

The geographic gossip algorithm uses location information to route packets on long
paths in the network. One natural extension of the algorithm is to allow all the nodes
on the routed path to be averaged jointly. This can be easily performed by aggregating the sum and the hop length while routing.  As long as the information of the average can be routed back on the same path, all the intermediate nodes can replace their estimates with updated value.
This modified algorithm is called \emph{geographic gossip with path averaging}. It was recently shown~\cite{BDTV_Allerton07} that this algorithm converges much faster, requiring only $\Theta (\sqrt{n}) $ gossip interactions and $\Theta (n \log \epsilon^{-1})$ messages, which is clearly minimal.


A related distributed algorithm was introduced by Savas et al.~\cite{DBLP:conf/dcoss/SavasAS06}, using multiple random walks
that merge in the network. The proposed algorithm does not require any  
location information
and uses the minimal number of messages,  $\Theta(n\log n)$,  to average  
on grid topologies with high probability. The
coalescence of information reduces the number of nodes that update  
information, resulting in optimal communication requirements  but also
less fault tolerance.  In most gossip algorithms all nodes keep  
updating their information which, as we discuss in the next  
section, adds robustness with respect to changes to the network and noise  
in communications.

Finally, we note the recent development of schemes that exploit the  
broadcast nature of wireless communications in order to accelerate  
gossip rates of convergence \cite{aysal-broadcast,ustebay08}, either  
by having all neighbors that overhear a transmission execute a local  
update, or by having nodes eavesdrop on their neighbors' communication  
and then using this information to strategically select which neighbor  
to gossip with next.  The next section discusses issues arising when  
gossiping specifically over wireless networks.

\section{Rate Limitations in Gossip Algorithms}
\label{sec:wireless}

Rate limitations are relevant due to the bandwidth restrictions and the power limitations of nodes.  Finite transmission rates imply that nodes learn of their neighbors' states with finite precision; if the distortion is measured by the MSE, then it is well established that the operational distortion rate function is exponentially decaying with the number of bits \cite{Ortega98}, which implies that the precision doubles for each additional bit of representation.
 For example, in an AWGN channel with path loss inversely proportional to the distance squared, $r^2$, the rate $R$ needs to be below the capacity bound $R<C= 1 /2 \log(1+\gamma r^{-2})$.  Then, at a fixed power budget, every bit of additional precision requires approximately shrinking the range by half; \emph{i.e.}, fixing $\gamma$, the channel capacity increases as the inter-node distance decreases. For a uniform network deployment, this would reduce the size of each node's neighborhood by about 75\%, decreasing the network connectivity and therefore the convergence speed.
This simple argument illustrates the importance of understanding if the performance of gossip algorithms degrades gracefully as the communication rate of each link decreases.

Before summarizing the key findings of selected literature on the subject of average consensus under communication constraints, we explain why some papers care about this issue and some do not.

\subsection{Are Rate Constraints Significant?}

In most sensor network architectures today, the overhead of packet headers and reliable communication is so great that using a few bytes to encode the gossip state variables exchanged leads to negligible additional cost while practically giving a precision that can be seen as infinite.  Moreover, we can ignore bit errors in transmissions, which very rarely go undetected thanks to CRC bits. It is natural to ask: \emph{why should one bother studying rate constraints at all?}

One should bother because existing sensor network modems are optimized to transmit long messages, infrequently, to nearby neighbors, in order to promote spatial bandwidth reuse, and were not designed with decentralized iterative computation in mind.  Transmission rates are calculated amortizing the overhead of establishing the link over the duration of very long transmission sessions.

Optimally encoding for computation in general (and for gossiping in particular) is an open problem; very few have treated the subject of communication for computation in an information theoretic sense (see, e.g., \cite{Orlitsky90,gastpar}) and consensus gossiping is nearly absent in the landscape of network information theory.  This is not an accident. Broken up in parts, consensus gossip contains the elements of complex classical problems in information theory, such as multi-terminal source coding, the two-way channel, the feedback channel, the multiple access of correlated sources and the relay channel \cite{Cover1991}; this is a frightening collection of open questions.
However, as the number of possible applications of consensus gossip primitives expands, designing source and channel encoders to solve precisely this class of problems more efficiently, even though perhaps not optimally,  is a worthy task. Desired features are efficiency in exchanging frequently, and possibly in an optimal order, few correlated bits, and exchanging with
nodes that are (at least occasionally) very far, to promote rapid diffusion. Such forms of communications are very important in sensor networks
and network control.

Even if fundamental limits are hard to derive,  there are several heuristics that have been applied to the problem to yield some achievable bound. Numerous papers have studied the effects of intermittent or lossy links in the context of gossip algorithms (i.i.d.~and correlated models, symmetric and asymmetric)~\cite{tsp07-K-M,Mesbahi,Bucklew,Bamieh-patterson,ChaiWu,Porfiri,Jakovetic-Moura,Jadbabai,karmoura-randomtopologynoise,Nedic-Ozdaglar-Parrilo}.  In these models, lossy links correspond to masking some edges from the topology at each iteration, and, as we have seen above, the topology directly affects the convergence rate.  Interestingly, a common thread running through all of the work in this area is that so long as the network remains connected on average, convergence of gossip algorithms is not affected by lossy or intermittent links, and convergence speeds degrade gracefully.

Another aspect that has been widely studied is that of source coding for average consensus, and is the one that we consider next in Section \ref{sec.quantized-consensus}.  It is fair to say that, particularly in wireless networks, the problem of channel coding is essentially open, as we will discuss in section \ref{sec.gossip-channel-coding}.

\subsection{Quantized consensus}\label{sec.quantized-consensus}

Quantization maps the state variable exchanged $x_j(t)$ onto codes that correspond to discrete points
$Q_{t,j}(x_j(t))=q_j(t) \in {\cal Q}_{t,j}\subset \mathbb{R}$.
%
The set ${\cal Q}_{t,j}$
is referred to as the {\it code} used at time $t$ by node $j$;
 the  points $q_j(t)$ are used to generate an approximation $\hat x_j(t)$
of the state $x_j(t) \in \mathbb{R}$ that each node needs to transmit; the quantizer {\it rate}, in bits, is  $R_{t,j}=\log_2|{\cal Q}_{t,j}|$, where $|{\cal A}|$ is the cardinality of the set ${\cal A}$. Clearly, under the constraints specified previously on the network update matrix $W(t)$, the consensus states, $\{c \ones : c\in {\mathbb R}\}$, are fixed points. The evolution of the nodes' quantized states
is that of an automaton; under asynchronous random exchanges, the network state forms a Markov chain with $\prod_{j=1}^n
|{\cal Q}_{t,j}|$ possible states and consensus states $\{c \ones : c\in {\mathbb R}\}$ that are {\it absorbing states}.
The cumulative number of bits that quantized consensus diffuses throughout the network asymptotically is:
\begin{equation}\label{eq.Rtot}
R^{\infty}_{tot}=\sum_{t=1}^{\infty} R_{t,tot}=\sum_{t=1}^{\infty}\sum_{j=1}^nR_{t,j}.
\end{equation}

\begin{figure}
\centering
  (a)\includegraphics[width=.7\columnwidth]{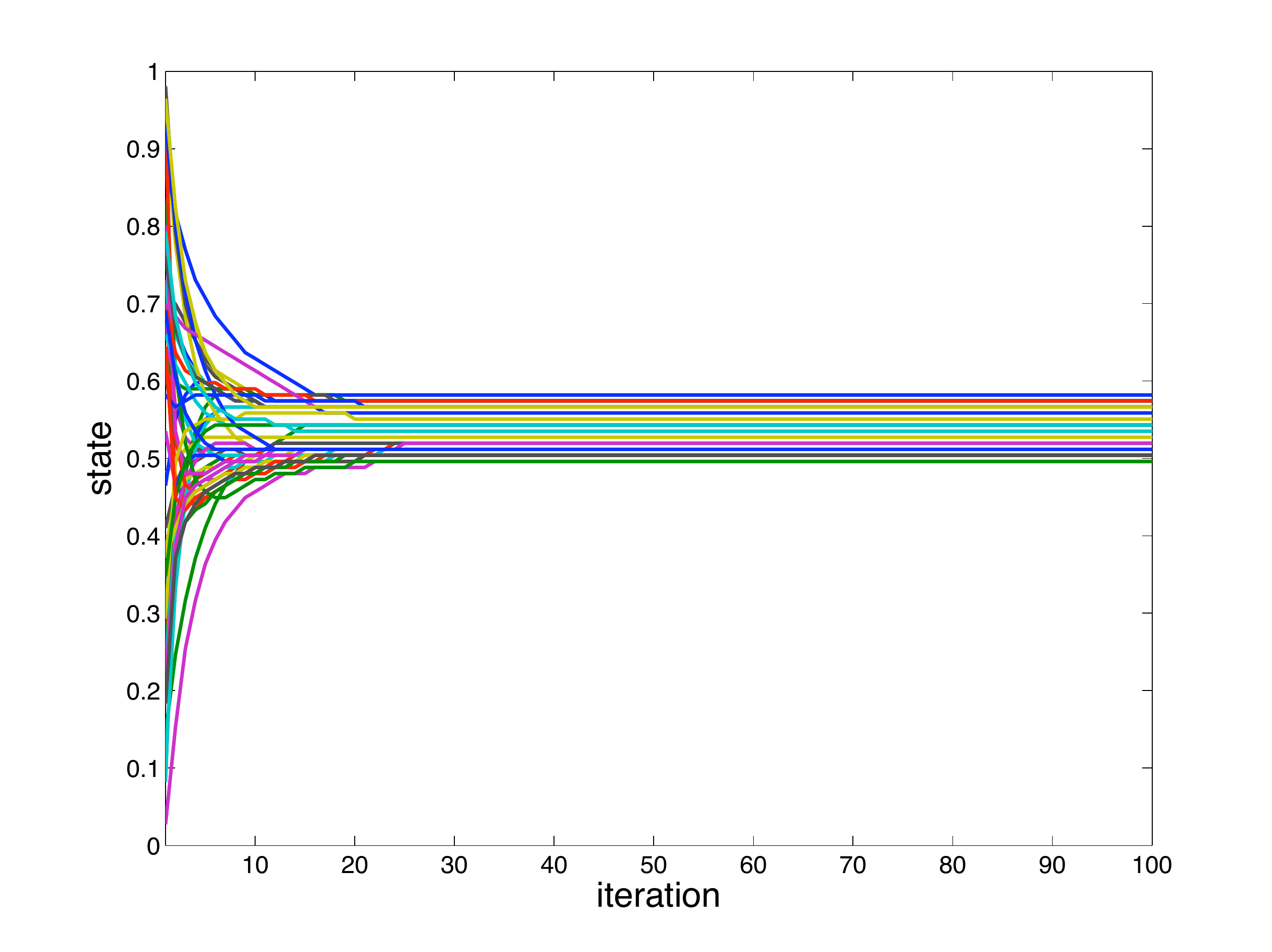}\\
  (b)\includegraphics[width=.7\columnwidth]{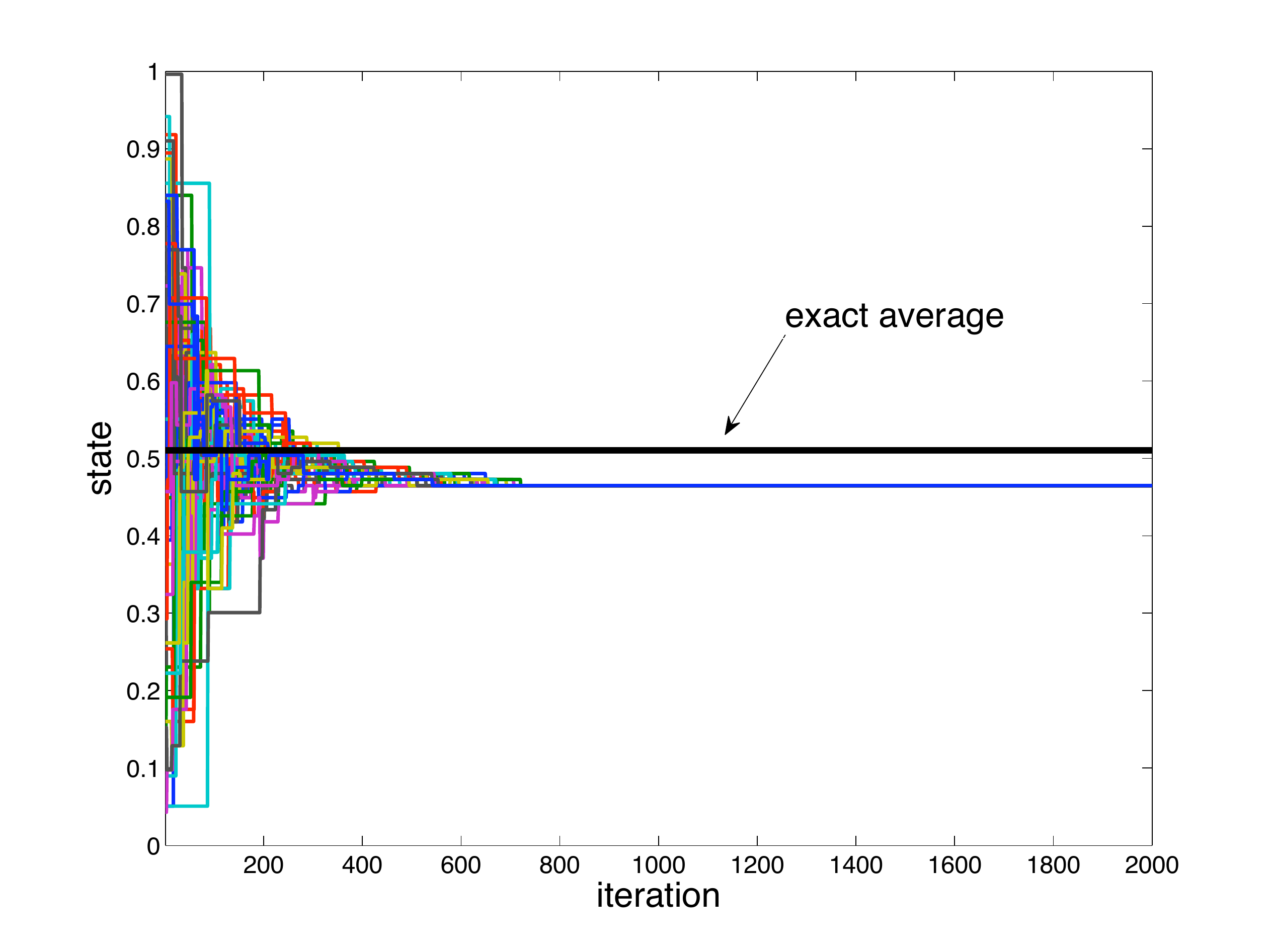}\\
 \center \caption{ Quantized consensus over a random geometric graph with $n=50$ nodes transmission radius $r=.3$ and initial states $\in [0,1]$ with uniform quantization with 128 quantization levels. Synchronous updates (a) and pairwise exchange (b).
}\label{fig.q-consensus}
\end{figure}

The first simple question is: for a fixed uniform quantizer with step-size $\Delta$, i.e.,
$$ \hat x_j(t)={\rm uni}_{\Delta}(x_j(t))={\rm arg}\!\min_{q\in {\cal Q}}|x_j(t)-q|, $$
where  ${\cal Q}=\{0,\pm \Delta, \pm 2 \Delta, \ldots, \pm (2^{R-1}-1)\Delta \}$, do the states $x(t)$ always converge (in a probabilistic sense) to the fixed points $c \ones$? The second is: what is the distortion $d\left(\lim_{k\rightarrow \infty}x(t), \frac 1 n \sum_{i=1}^n x_i(0) \ones\right)$ due to limited $R_{k,i}$ or a total budget $R_{tot}$?
Fig.~\ref{fig.q-consensus} illustrates the basic answers through numerical simulation.  Interestingly, with a synchronous gossip update, quantization introduces new fixed points other than consensus (Fig. \ref{fig.q-consensus}(a)), and asynchronous gossiping in general reaches consensus, but without guarantees on the location of outcome (Fig. \ref{fig.q-consensus}(b)).

Kashyap \emph{et~al.} \cite{Kashyap2007} first considered a fixed code quantized consensus algorithm, which preserves the network average at every iteration.
In their paper, the authors draw an analogy between quantization and load balancing among processors, which naturally comes with an integer constraint since the total number of tasks are finite and divisible only by integers (see, e.g., \cite{cybenko89,Subramanian94,Rabani98}). Distributed policies to attain a balance among loads were previously proposed in \cite{Aiello93,Ghosh96}. 
Assuming that the average can be written as $\frac 1 n \sum_{j=1}^n x_j(0)= \frac S n$ and denoting $L \triangleq S~ {\rm mod}~ n$,
under these updates in \cite{Kashyap2007} it is proven that any algorithm meeting the aforementioned conditions
makes every node converge to either $L$ or $L+1$, thereby approximating the average. The random gossip algorithm analyzed in \cite{Murray09} leads to a similar result, where the final consensus state differs at most by one bin from the true average; the same authors discuss bounds on the rate of convergence in \cite{Lavaei09}.
In these protocols the agents will be uncertain on what interval contains the actual average: the nodes whose final value is $L$ will conclude that the average is in $[L-1,L+1]$ and those who end with $L+1$ will think that the average is in $[L,L+2]$.
Benezit et al.~\cite{Benezit09} proposed a slight modification of the policy, considering a fixed rate class of quantization strategies that are based on {\it voting},  requiring only 2 bits of memory per agent and attaining a consensus on the interval that contains
the actual average.

To overcome the fact that not all nodes end up having the same quantized value, a simple variant on the quantized consensus problem that guarantees almost sure convergence to a unique consensus point was proposed concurrently in \cite{karmoura-quantized} and \cite{Aysal07}. The basic idea is to dither the state variables by adding a uniform random variable $u\sim {\cal U}(-\frac{\Delta} 2, \frac{\Delta} 2)$ prior to quantizing the states, i.e., $ \hat x_i(t)={\rm uni}_{\Delta}(x_i(t)+u)$.  This modest change enables gossip to converge to a consensus \textit{almost surely}, as shown in ~\cite{Aysal08}. This guarantees that the nodes will make exactly the same decision.  However, the algorithm can deviate more from the actual average than the quantized consensus policies considered in \cite{Kashyap2007}.
\begin{figure}
\centering
  \includegraphics[width=.7\columnwidth]{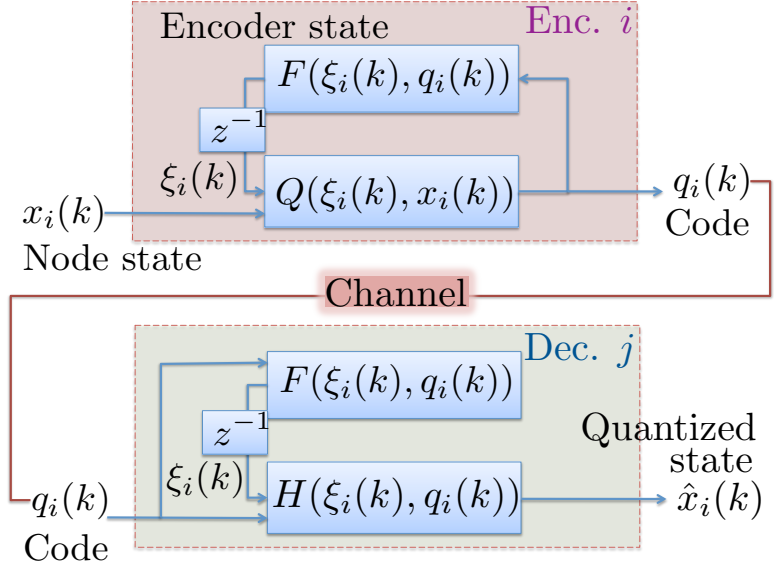}\\
 \center \caption{Quantized consensus  Node $i$ Encoder and Node $j$ Decoder, with memory.}\label{fig.carli}
\end{figure}
The advantage of using a fixed code is the low complexity, but with relatively modest additional cost, the performance can considerably improve.

Carli et al. in \cite{Carli08} noticed that the issue of quantizing for consensus averaging has analogies with the problem of stabilizing a system using quantized feedback \cite{Elia01}, which amounts to partitioning the state-space into sets whose points can be mapped to an identical feedback control signal.  Hence, the authors resorted to control theoretic tools to infer effective strategies for quantization. In particular, instead of using a static mapping, they model the quantizer $Q_t(x_i(t))$ at each node $i$ as a dynamical system with internal state $\xi_i(t)$, which is coupled with the consensus update through a quantized {\it error} variable $q_i(t)$ (see Fig. \ref{fig.carli}).
They study two particular strategies.  They refer to the first as the {\it zoom in - zoom out uniform coder/decoder} where they adaptively quantize the state as follows.  The node states are defined as
\begin{eqnarray}
 \xi_i(t)&=& (\hat x_{-1,i}(t),f_i(t)).  \label{eq.zoom1}
\end{eqnarray}
The quantized feedback and its update are
\begin{eqnarray}
 \hat x_i(t)&=& \hat x_{-1,i}(t+1)= \hat x_{-1,i}(t)+f_i(t)q_i(t); \label{eq.zoom2}\\
q_i(t)&=&{\rm uni}_{\Delta}\left(\frac{x_i(t)- \hat x_{-1,i}(t)}{f_i(t)}\right), \label{eq.zoom3}
\end{eqnarray}
which is basically a differential encoding, and $f_i(t)$ is the stepsize, updated according to
\begin{eqnarray}
f_i(t+1)&=&\left\{
\begin{array}{l l}
 k_{in}f_i(t) & {\rm if}~~|q_i(t)|<1     \\
 k_{out}f_i(t) &   {\rm if}~~|q_i(t)|=1
\end{array}\right., \label{eq.zoom4}
\end{eqnarray}
which allows the encoder to adaptively zoom-in and out, depending on the range of $q_i(t)$.
The second strategy has the same node states but uses a logarithmic quantizer,
\begin{eqnarray}\label{eq.logarithmic}
 \hat x_i(t)&=& \xi_i(t+1)=  \xi_i(t)+q_i(t);\\
q_i(t)&=&{\log}_{\delta}(x_i(t)- \xi_i(t)),
\end{eqnarray}
where the logarithmic quantization amounts to:
\begin{eqnarray}
q_i(t)&=&{\rm sign}(x_i(t)- \xi_i(t))\left(\frac{1+\delta}{1-\delta} \right)^{\ell_i(t)}\\
\ell_i(t)&: & \frac 1 {1-\delta}\leq |x_i(t)- \xi_i(t)|\left(\frac{1+\delta}{1-\delta} \right)^{-\ell_i(t)}\!\!\!\leq \frac{1}{1+\delta}. \nonumber
\end{eqnarray}
In \cite{Carli08} numerical results are provided for the convergence of the {\it zoom-in/out} quantizer, while the properties of the logarithmic quantizer are studied analytically. Remarkably, the authors prove that if the state average is preserved and if $0<\delta<\frac{1+\lambda_{\min}(W)}{3-\lambda_{\min}(z)}$, then the network reaches asymptotically exactly the same state as the un-quantized average consensus.  In other words, for all $i$, $\lim_{k\rightarrow \infty}x_i(t)= \frac 1 n \sum_{i=1}^n x_i(0)$.
One needs to observe that the logarithmic quantizer replaces state values in an uncountable set ${\mathbb R}$ with discrete countable outputs $\ell \in {\mathbb N}$, in the most efficient way \cite{Elia01}, but there are still infinite many such sets; in other words, the logarithmic quantizer has unlimited range and therefore $R_{t,j}=\infty$. Hence,
in practice, one will have to accept a penalty in accuracy when its range is limited.

\begin{figure}
\centering
  \includegraphics[width=.7\columnwidth]{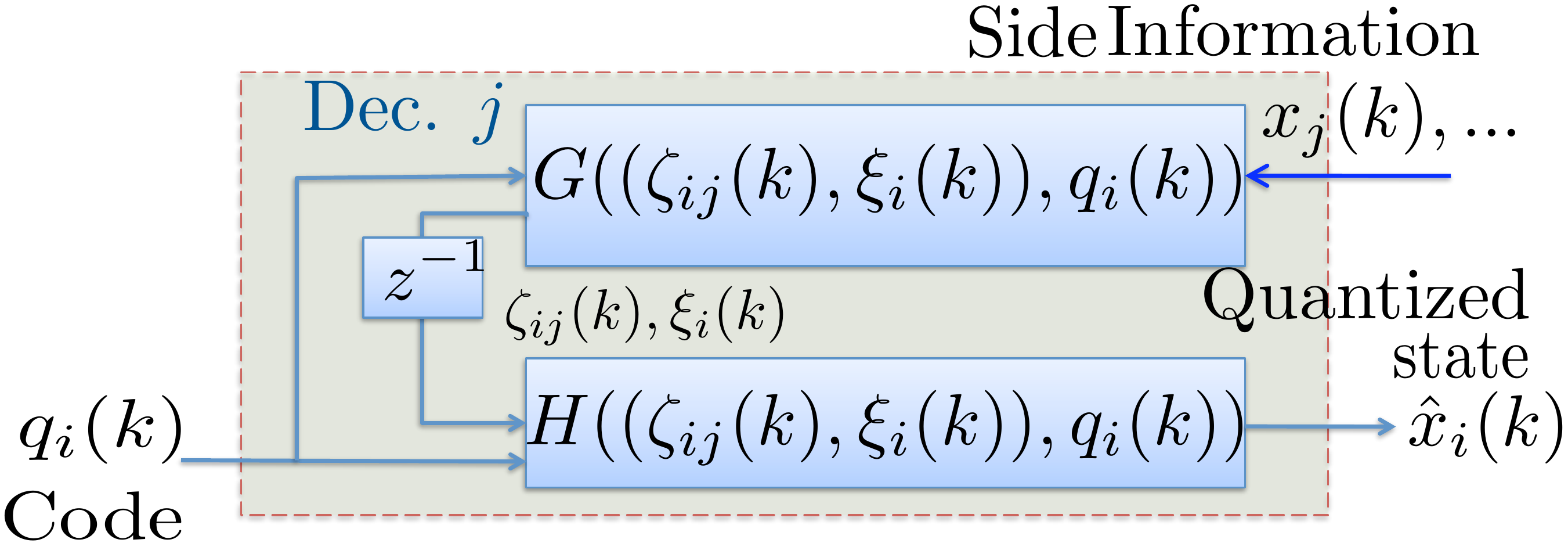}\\
 \center \caption{ Node $j$ Decoder, with memory and side information.}\label{fig.ercan}
\end{figure}

The vast signal processing literature on sampling and quantization can obviously be applied to the consensus problem as well to find heuristics.
It is not hard to recognize that the quantizers analyzed in \cite{Carli08} are equivalent to predictive quantizers. Noting that the states are both temporally and spatially correlated, it is clear that encoding using the side information that is available at both transmitter and receiver can yield improved performance and lower cost; this is the tenet of the work in \cite{yildizIPSN07}, \cite{yildizSPT08}, which analyzed a more general class of quantizers. They can be captured in a similar framework as that of \cite{Carli08} by adding an auxiliary state variable $\zeta_{ij}(t)$, which affects the state of the decoder only (see the decoder in Fig. \ref{fig.ercan}).
The idea is similar, since
$\hat x_{-1,i}(t+1)$ in \eqref{eq.zoom2} is replaced in \cite{yildizSPT08} by the optimum linear minimum mean-squared error prediction, performed using $k$ previous states $\hat x_{-1,i}(t)=\sum_{l=1}^k a_{i,k}(l)\hat x_i(t-l)$. Similarly, the receiver state is introduced to utilize the idea of {\it coding with side information} \cite{WynerZiv76}, where the side information about $x_i(t)$ that is available at, say, receiver $j$ consists of the receiver's present state $x_j(t)$, as well as possibly its own past states and those of neighbors in communication with node $j$.
The decoder augmented state $(\zeta_{ij}(t),\xi_i(t))$ in  \cite{yildizSPT08} is useful to reap the benefits of the refinement in the prediction of $x_i(t)$ that the decoder can obtain using its own side information. This prediction $\hat x_{-1,ij}(t)$ can more closely approximate the true state $x_i(t)$ compared to the transmitter $\hat x_{-1,i}(t)$ and this, in turn, means that \eqref{eq.zoom3} can be replaced by a nested  quantizer, such as for example the nested lattice quantizers in \cite{Zamir02}.
 In practice, to keep the complexity at bay,
one can use a static nested lattice quantizer at the transmitter without any memory, while using the current local state as the $j$-node decoder state, i.e., $\zeta_{ij}(t)=x_j(t)$.
The main analytical result in \cite{yildizSPT08} is the conclusion that, even with the lowest complexity (i.e. prediction memory $k=1$ only or $\zeta_{ij}(t)=x_j(t)$ and no memory) one needs finite $R_{tot}^{\infty}<\infty$ to guarantee that the network will reach consensus with a bounded error
$d\left(\lim_{k\rightarrow \infty}x(t), \frac 1 n \sum_{i=1}^n x_i(0) \ones \right)\leq D_{tot}^{\infty}$ that decreases as a function of $R_{tot}^{\infty}$. This is useful to establish since one may argue that, as long as the network is finite, flooding each value from each node, rather than gossiping, would require a total cost in term of transmission bits that is finite, which can also be reduced via optimal joint source and network coding methods. It is meaningful to ask if gossiping can also lead to a similar rate-distortion tradeoff and the result in \cite{yildizSPT08} suggests that this is, indeed, the case.

Recent work has begun to investigate information-theoretic performance bounds for gossip.  These bounds characterize the rate-distortion tradeoff either (i) as a function of the underlying network topology assuming that each link has a finite capacity~\cite{ayaso}, or (ii) as a function of the rate of the information source providing new measurements to each sensor~\cite{su-elgamal}.

\subsection{Wireless channel coding for average consensus}\label{sec.gossip-channel-coding}

Quantization provides a source code, but equally important is the channel code that is paired with it.  First, using the separation of source and channel coding in wireless networks is not optimal in general.  Second, and more intuitively, in a wireless network there are a variety of rates that can be achieved with a variety of nodes under different traffic conditions.  The two key elements that determine what communications can take place are scheduling and channel coding.  Theoretically, there is no fixed-range communication; any range can be reached albeit with lower capacity.  Also, there is no such thing as a collision; rather, there is a tradeoff between the rate that multiple users can simultaneously access the channel.

The computational codes proposed in \cite{nazer-gastpar} aim to strike a near-optimal trade-off for each gossip iteration, by utilizing the additive noise multiple access channel as a tool for directly computing the average of the neighborhood. The idea advocated by the authors echoes their previous work \cite{gastpar}: nodes send lattice codes that, when added through the channel, result in a lattice point that encodes a specific algebraic sum of the inputs. Owing to the algebraic structure of the channel codes and the linearity of the channel, each recipient decodes directly the linear combination of the neighbors' states, which provides a new estimate of the network average when added to the local state. The only drawbacks of this approach is that 1) it requires channel state information at the transmitter, and 2) that only one recipient can be targeted at the time.  The scenario considered is closer to that in \cite{su-elgamal}, since a stream of data needs to be averaged, and a finite round is dedicated to each input.  The key result proven is that the number of rounds of gossip grows as $O(\log n^2 /r^2)$ where $r$ is the radius of the neighborhood.

\section{Sensor Network Applications of Gossip}
\label{sec:apps}

This section illustrates how gossip algorithms can be applied to solve representative problems in wireless sensor networks.
Of course, gossip algorithms are not suited for all distributed signal processing tasks.  They have proven useful, so far, for problems that involve computing functions that are linear combinations of data or statistics at each node.  Two straightforward applications arise from distributed inference and distributed detection.  When sensors make conditionally independent observations, the log-likelihood function conditioned on a hypothesis $H_j$ is simply the sum of local log-likelihood functions, $\sum_{i=1}^n \log p(x_i | H_j)$, and so gossip can be used for distributed detection (see also \cite{Kar07a,saligrama06}).  Similarly, if sensor readings can be modeled as i.i.d.~Gaussian with unknown mean, distributed inference of the mean boils down to computing the average of the sensor measurements, and again gossip can be applied.  Early papers that made a broader connection are those of Saligrama et al.~\cite{saligrama06}, and Moallemi and Van Roy~\cite{Moallemi06}, which both discuss connections between gossip algorithms and belief propagation.

Below we consider three additional example applications.  Section~\ref{subsec:robustEstimation} describes a gossip algorithm for distributed linear parameter estimation that uses stochastic approximation to overcome quantization noise effects.  Sections~\ref{subsec:sourceLoc} and \ref{subsec:fieldEst} illustrate how gossip can be used for distributed source localization and distributed compression, respectively.  We also note that gossip algorithms have recently been applied to problems in camera networks for distributed pose estimation~\cite{Tron08,Jorstad08}.

\subsection{Robust Gossip for Distributed Linear Parameter Estimation}
\label{subsec:robustEstimation}

The present section focuses on \emph{robust} gossiping for distributed linear parameter estimation of a vector of parameters with low-dimensional observations at each sensor. We describe the common assumptions on
sensing, the network topology, and the
gossiping protocols. Although we focus on estimation, the formulation is quite general and applies to
many inference problems, including distributed detection and distributed localization.

\subsubsection{Sensing/Observation Model} Let $\theta \in \mathbb{R}^{m \times 1}$ be an $m$-dimensional parameter that is to be estimated by a network of~$n$ sensors. We refer to~$\theta$ as a parameter, although it is a vector of~$m$ parameters. For definiteness we assume the following observation model for the $i$-th sensor:
\begin{equation}
\label{obsmod}
z_{i}(t)= H_{i} \theta + w_i(t)
\end{equation}
where: \begin{inparaenum}[] \item $\left\{z_{i}(t)\in\mathbb{R}^{m_{i} \times 1}\right\}_{t\geq 0}$ is the i.i.d. observation sequence for the
$i$-th sensor; \item $\left\{w_{i}(t)\right\}_{t\geq 0}$ is a zero-mean i.i.d.~noise sequence of bounded variance.
\end{inparaenum}
For most practical sensor network applications, each sensor observes only a subset of~$m_i$ of the components of $\theta$, with $m_{i} \ll m$. Under such conditions, in isolation, each sensor can estimate at most only a part of the parameter. Since we are interested in obtaining a consistent estimate of the entire parameter $\theta$ at each sensor, we need some type of observability condition. We assume the matrix
 \begin{equation}
 \label{obs_rank} \sum_{i=1}^{n} H_{i}^{T} H_{i}
 \end{equation}
 is full rank. Note that the invertibility is even required by a centralized estimator
 (one which has access to data from all sensors at all time) to get a consistent
 estimate of $\theta$. It turns out that, under reasonable assumptions on
 the network connectivity, this necessary condition for centralized observability is sufficient for
 distributed observability, i.e., for each sensor to obtain a consistent estimate of
 $\theta$. It is not necessary to restrict to time-invariant observation
 matrices and the $H_{i}$s can be random time-varying~\cite{Asilomar08-K-M},
 as would be required in most regression based analyses. In general, the observations need
 not come from a linear statistical model and may be distributions parameterized by
 $\theta$. The distributed observability would then correspond to
 asymptotic distinguishability of the collection of these distributions over the network.  A generic formulation in such a setting requires
 the notion of separably estimable observation models (see~\cite{kar-moura-ramanan-IT-2008}).

 An equivalent formulation of the estimation problem in the setting considered above, comes from the distributed \emph{least mean square} (LMS) adaptive filtering
 framework~\cite{Sayed-LMS,Stankovic-parameter,Giannakis-LMS}. The objective here is slightly
 different. While we are interested in consistent estimates of the entire parameter at each sensor,
  the LMS formulations require, in a distributed way, to adapt to the environment to produce a desired
  response at each sensor, and the observability issue is not of primary importance.
  A generic framework for distributed estimation, both in the static parameter case and when
   the parameter is non-stationary, is addressed in~\cite{Nedic-parameter}. An important aspect
   of algorithm design in these cases is the choice of the inter-sensor weight sequence for fusing
   data or estimates. In the static parameter case, where the objective is to drive all the sensors
   to the true parameter value, the weight sequence necessarily decays over time to overcome the
   accumulation of observation and other forms of noises, whereas, in the dynamic parameter
   estimation case, it is required that the weight sequence remains bounded away from zero,
   so that the algorithm possesses tracking abilities.
 We direct the reader to the recent article~\cite{Sundarbook} for a discussion along these
 lines. In the dynamic case, we also suggest the significant literature on distributed
 Kalman filtering (see,
 e.g.,~\cite{Olfati-Kalman,Scaglione-Kalman,Khan-Moura, Giannakis-Kalman, Schenato-Kalman}
 and the references therein), where the objective is not consensus seeking among the local estimates,
 but, in general, optimizing fusion strategies to minimize the mean-squared error at each
 sensor.

 It is important to note here that average consensus is a specific case of a distributed parameter estimation model, where each sensor initially takes a single measurement, and sensing of the field thereafter is not required for the duration of the gossip algorithm.  Several distributed inference protocols (for example,~\cite{Mesbahi-parameter,Giannakis-est,tsp06-K-A-M}) are based on this approach, where either the
sensors take a single snapshot of the field at the start and then
initiate distributed consensus protocols (or more generally
distributed optimization, as in~\cite{Giannakis-est}) to fuse the
initial estimates, or the observation rate of the sensors is
assumed to be much slower than the inter-sensor communicate rate,
thus permitting a separation of the two time-scales.

\subsubsection{Distributed linear parameter estimation}
 We now briefly discuss distributed parameter estimation in the linear observation model~(\ref{obsmod}). Starting from an initial deterministic estimate of the parameters (the initial states may be random, we assume deterministic for notational simplicity), $x_{i}(0)\in \mathbb{R}^{m\times 1}$, each sensor generates, by a distributed iterative algorithm, a sequence of estimates, $\left\{x_{i}(t)\right\}_{t\geq 0}$. To simplify the discussion in this section, we assume a synchronous update model where all nodes exchange information and update their local estimates at each iteration.  The parameter estimate $x_{i}(t+1)$ at the $i$-th sensor at time $t+1$ is a function of: \begin{inparaenum}[1)] \item its previous estimate; \item the communicated quantized estimates at time~$t$ of its neighboring sensors; and \item the new observation $z_{i}(t)$.
\end{inparaenum}
 The data is subtractively dithered quantized, i.e., there exists a vector quantizer~$Q(.)$
 and a family, $\left\{\nu_{ij}^{l}(t)\right\}$, of i.i.d.~uniformly distributed random variables on
 $[-\Delta/2,\Delta/2)$ such that the quantized data received by the $i$-th sensor from the
 $j$-th sensor at time~$t$ is $Q(x_{j}(t) + \nu_{ij}(t))$,
 where $\nu_{ij}(t)=[\nu_{ij}^{1}(t), \cdots,\nu_{ij}^{m}(t)]^{T}$. It then
 follows  that the quantization error,
 $\varepsilon_{ij}(t) \in \mathbb{R}^{m\times 1}$, 
is a random vector, whose components are i.i.d.~uniform on $[-\Delta/2, \Delta/2)$ and independent of $x_{j}(t)$.

\subsubsection{Stochastic approximation algorithm} Let $\mathcal{N}_i(t)$ denote the neighbors of node $i$ at iteration $t$; that is, $j \in \mathcal{N}_i(t)$ if $i$ can receive a transmission from $j$ at time $t$.  In this manner, we allow the connectivity of the network to vary with time.  Based on the current state, $x_{i}(t)$, the quantized exchanged data $\left\{Q(x_{j}(t)+ \nu_{ij}(t))\right\}_{j\in\mathcal{N}_{i}(t)}$, and the observation~$z_{i}(t)$, the updated estimate at node $i$ is
\begin{eqnarray}
\label{algRE}
x_{i}(t+1)&=& x_{i}(t)-\alpha(t)
\left[b\sum_{j\in\mathcal{N}_{i}(t)}( x_{i}(t)\right.\\
\nonumber
&-&
Q(x_{j}(t)+ \nu_{ij}(t)))
- H_{i}^{T}\left( z_{i}(t)\right.\\
\nonumber
&-&
\left.
H_{i} x_i(t)\right)\Big]
\end{eqnarray}
In~(\ref{algRE}),  $b>0$ is a constant and $\left\{\alpha(t)\right\}_{t\geq0}$ is a sequence of weights satisfying the persistence condition\footnote{We need the $\alpha(t)$ to sum to infinity, so that the algorithm `persists' and does not stop; on the other hand, the $\alpha$ sequence should be square summable to prevent the build up of noise over time.}:
\begin{equation}
\label{per_cond}\alpha(t)\geq 0,~~\sum_{t}\alpha(t)=\infty,~~\sum_{t}\alpha^{2}(t)<\infty
\end{equation}
 Algorithm~(\ref{algRE}) is distributed because for sensor~$n$ it involves only the data from the sensors in its neighborhood~$\mathcal{N}_i(t)$. 
 
The following result from~\cite{kar-moura-ramanan-IT-2008} characterizes the desired statistical properties of the distributed parameter estimation algorithm just described.  The flavor of these results is common to other stochastic approximation algorithms~\cite{Nevelson}.  First, we have a law of large numbers-like result which guarantees that the estimates at each node will converge to the true parameter estimates,
    \begin{equation}
\label{REas:1}
\mathbb{P}\left(\lim_{t\rightarrow\infty} x_{i}(t)
= \theta,~\forall i \right)=1
\end{equation}
If, in addition to the conditions mentioned above, the weight sequence is taken to be
\begin{equation}
\label{asynorm1} \alpha(t)=\frac{a}{t+1},
\end{equation}
for some constant $a>0$, we also obtain a central limit theorem-like result, describing the distribution of estimation error over time.  Specifically, for $a$ sufficiently large, we have that the error, $$\sqrt{t} \big(x(t) - \ones \otimes \theta\big)$$ converges in distribution to a zero-mean multivariate normal with covariance matrix that depends on the observation matrices, the quantization parameters, the variance of the measurement noise, $w_i(t)$, and the constants $a$ and $b$.  The two most common techniques for analyzing stochastic approximation algorithms are stochastic Lyapunov functions and the ordinary differential equations method~\cite{Nevelson}.  For the distributed estimation algorithm~\eqref{algRE}, the results just mentioned can be derived using the Lyapunov approach~\cite{Huang07,kar-moura-ramanan-IT-2008}.

\begin{figure*}
\centering %
\subfigure[][]{\includegraphics[height=2.1in, width=2.5in ]{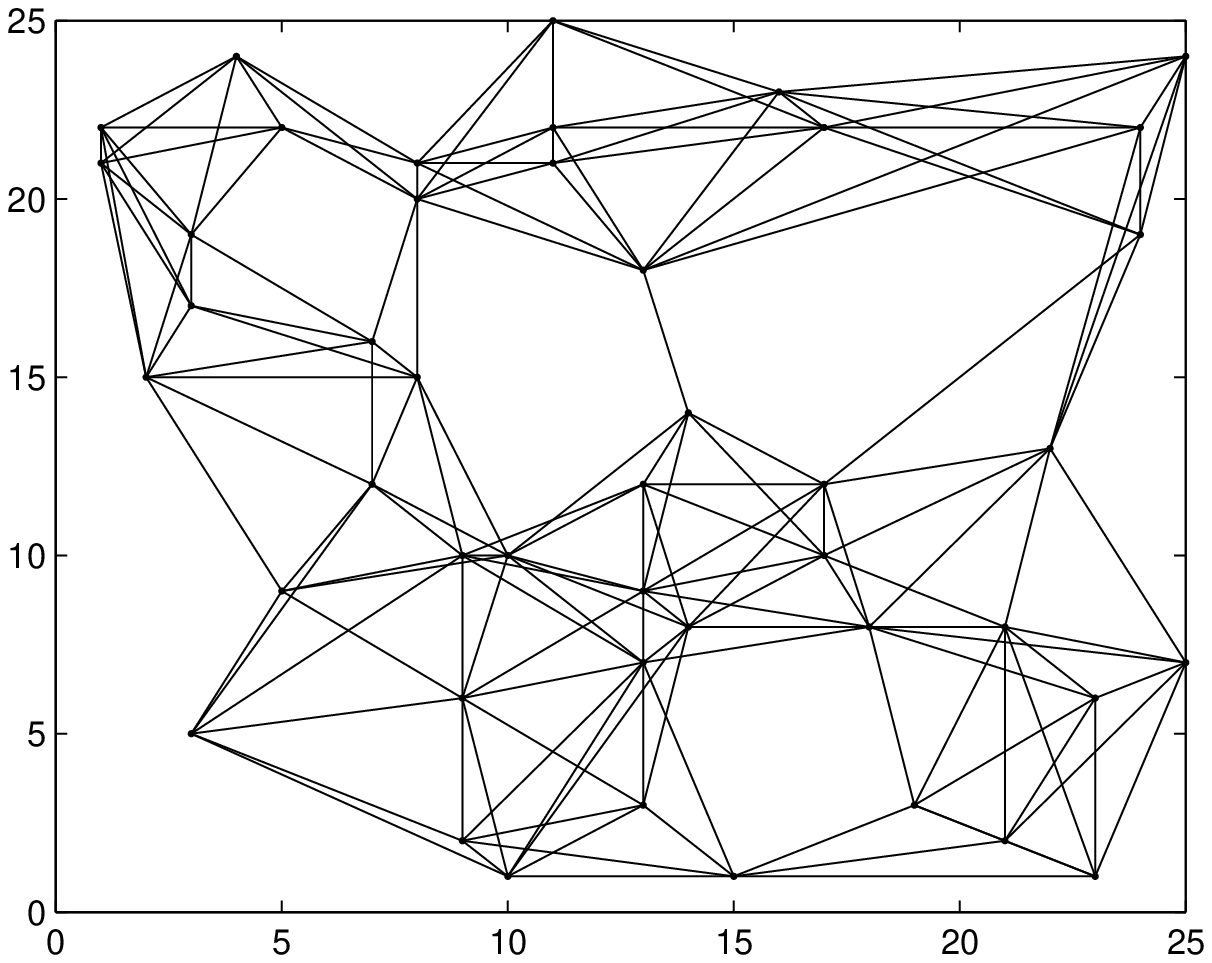}}
\subfigure[][]{\includegraphics[height=2.1in, width=2.5in ]{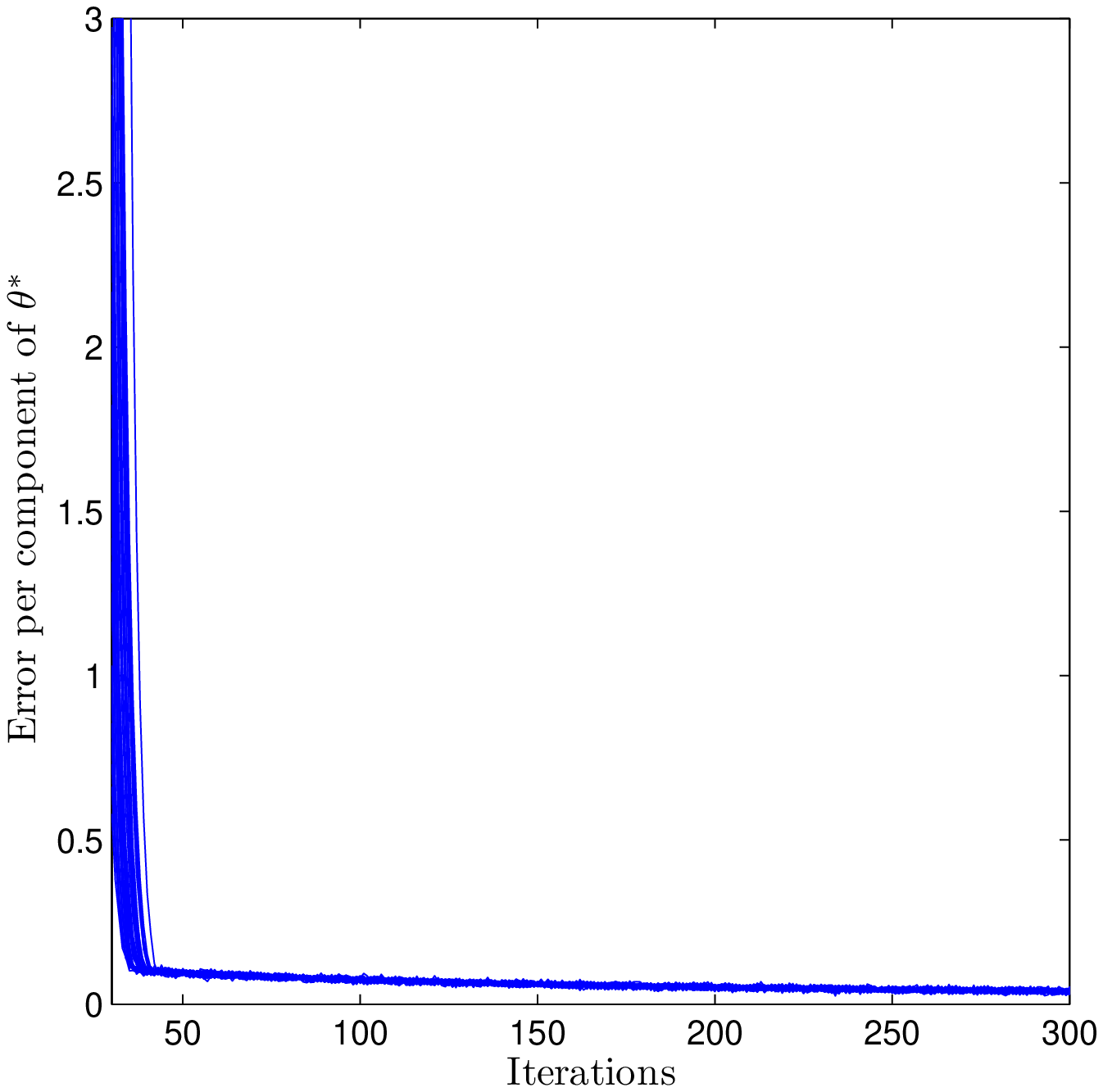}}
\caption{Illustration of distributed linear parameter estimation. (a)
Example network deployment of 45 nodes.  (b) Convergence of normalized estimation error
at each sensor.}
\label{fig:LU}
\end{figure*}

Performance analysis of the algorithm for an example network is illustrated in Figure~\ref{fig:LU}.  An example network of $n=45$ sensors are deployed randomly on a $25\times 25$ grid, where sensors communicate in a fixed radius and are further constrained to have a maximum of $6$ neighbors per node. The true parameter $\mathbf{\theta}^{\ast}\in\mathbb{R}^{45}$.  Each node is associated with a single component of $\mathbf{\theta}^{\ast}$. For the experiment, each component of $\mathbf{\theta}^{\ast}$ is generated by an instantiation of a zero mean Gaussian random variable of variance 25. (Note, the parameter $\mathbf{\theta}^{\ast}$ here has a physical significance and may represent the state of the field to be estimated. In this example, the field is assumed to be white, stationary and hence each sample of the field has the same Gaussian distribution and independent of the others. More generally, the components of $\mathbf{\theta}^{\ast}$ may correspond to random field samples, as dictated by the sensor deployment, representing a discretization of the PDE governing the field.) Each sensor observes the corresponding field component in additive Gaussian noise. For example, sensor 1 observes $z_{1}(t)=\theta^{\ast}_{1}+w_{1}(t)$, where $w_{1}(t)\sim\mathcal{N}(0,1)$. Clearly, such a model satisfies the distributed observability condition
\begin{equation}
\label{fig_LU1}
G=\sum_{i}H_{i}^{T}H_{i}=I=G^{-1}
\end{equation}
(Note, here $H_{i}=\mathbf{e}_{i}^{T}$, where $\mathbf{e}_{i}$ is the standard unit vector with 1 at the $i$-th component and zeros elsewhere.)
Fig.~\ref{fig:LU}(a) shows the network topology, and Fig.~\ref{fig:LU}(b) shows the normalized error of each sensor plotted against the iteration index $t$ for an instantiation of the algorithm. The normalized error for the $i$-th sensor at time $t$ is given by the quantity $\left\|\mathbf{x}_{i}(t)-\mathbf{\theta}^{\ast}\right\|/45$, i.e., the estimation error normalized by the dimension of $\mathbf{\theta}^{\ast}$. We note that the errors converge to zero as established by the theoretical findings. The decrease is rapid at the beginning and slows down at $t$ increases. This is a standard property of stochastic approximation based algorithms and is attributed to the decreasing weight sequence $\alpha(t)$ required for convergence.

It is interesting to note that, although the individual sensors suffer from low rank observations of the true parameter, by collaborating, each of them can reconstruct the true parameter value. The asymptotic normality shows that the estimation error at each sensor decays as $1/\sqrt{t}$, the decay rate being similar to that of a centralized estimator having access to all the sensor observations at all times. The efficiency of the distributed estimator is measured in terms of its asymptotic variance, the lower limit being the Fisher information rate of the corresponding centralized estimator. As expected, because of the distributed nature of the protocol (information needs to disseminate across the entire network) and quantized (noisy) inter-sensor communication, the achieved asymptotic variance is larger than the centralized Fisher information rate. In the absence of quantization (perfect communication), it can be shown that the parameter $a$ in eqn.~(\ref{asynorm1}) can be designed appropriately so that the asymptotic variance of the decentralized estimator matches the centralized Fisher information rate, showing that the distributed estimator described above is efficient for the centralized estimation problem. An example of interest, with Gaussian observation noise is studied in~\cite{kar-moura-ramanan-IT-2008}, where it is shown that asymptotic variance attainable by the distributed algorithm is the same as that of the optimum (in the sense of Cram\'er-Rao) centralized estimator having access to all information simultaneously. This is an interesting result, as it holds irrespective of the network topology. Such a phenomenon is attributed to a time scale separation between the consensus potential and the innovation rate (rate of new information entering the network), when inter-sensor communication is unquantized (perfect) with possible link failures.

As noted before, the observation model need not be linear for distributed parameter estimation. In~\cite{kar-moura-ramanan-IT-2008} a large class of nonlinear observation models were considered and a notion of distributed nonlinear observability called separably estimable observable models introduced. Under the separably estimable condition, there exist local transforms under which the updates can be made linear. However, such a state transformation induces different time scales on the consensus potential and the innovation update, giving the algorithm a mixed time-scale behavior (see~\cite{kar-moura-ramanan-IT-2008},\cite{ICASSP09-K-M} for details.) This mixed time-scale behavior and the effect of \emph{biased} perturbations leads to the inapplicability of standard stochastic approximation techniques.

\subsection{Source Localization}
\label{subsec:sourceLoc}

A canonical problem, encompassing many of the challenges which commonly arise in wireless sensor network applications, is that of estimating the location of an energy-emitting source \cite{zhao04}.  Patwari et al.~\cite{patwari05} presents an excellent overview of the many approaches that have been developed for this problem.  The aim in this section is to illustrate how gossip algorithms can be used for source localization using received signal strength (RSS) measurements.

Let $\theta \in \Re^2$ denote the coordinates of the unknown source, and for $i=1,\dots,n$, let $y_i \in \Re^2$ denote the location of the $i$th sensor.  The RSS measurement at node $i$ is modeled as
\begin{equation}
f_i = \frac{\alpha}{\|y_i - \theta\|^\beta} + w_i,
\end{equation}
where $\alpha > 0$ is the signal strength emitted at the source, $\beta$ is the path-loss coefficient, and $w_j$ is additive white Gaussian noise.  Typical values of $\beta$ are between 2 and 4.  This model was validated experimentally in \cite{li03}.  Centralized maximum likelihood estimators for single and multiple-source scenarios based on this model are presented in \cite{sheng03} and \cite{sheng05}.  Because the maximum likelihood problem is, in general, non-linear and non-convex, it is challenging to solve in a decentralized fashion.  Distributed approaches based on finding a cyclic route through the network are presented in \cite{rabbat04,blatt06}.

An alternative approach, using gossip algorithms~\cite{rabbat05}, forms a location estimate $\widehat{\theta}$ by taking a linear combination of the sensor locations weighted by a function of their RSS measurement,
\begin{equation}
\widehat{\theta} = \frac{\sum_{i=1}^n y_i K(f_i)}{\sum_{i=1}^n K(f_i)}, \label{eqn:sourceLocEstimator}
\end{equation}
where $K : \Re^+ \rightarrow \Re^+$ is a monotone increasing function satisfying $K(0) = 0$ and $\lim_{f \rightarrow \infty} K(f) < \infty$.  Intuitively, nodes that are close to the source measure high RSS values, so their locations should be given more weight than nodes that are further away, which will measure lower RSS values.  Taking $K(f) = 1_{\{f \ge \gamma\}}$, where $\gamma > 0$ is a positive threshold, and where $1_{\{\cdot\}}$ is the indicator function, \eqref{eqn:sourceLocEstimator} reduces to
\begin{equation}
\widehat{\theta}_1 = \frac{\sum_{i=1}^n y_i 1_{\{\|y_i - \theta\| \le \gamma^{-1/\beta}\}}}{\sum_{i=1}^n 1_{\{\|y_i - \theta\| \le \gamma^{-1/\beta}\}}},
\end{equation}
which is simply the centroid of the locations of sensors that are no further than $\gamma^{-1/\beta}$ from the source.  In~\cite{rabbat05}, it was shown that this estimator benefits from some attractive properties.  First, if the sensor locations $y_i$ are modeled as uniform and random over the region being sensed, then $\widehat{\theta}_1$ is a consistent estimator, as the number of sensors grows.  It is interesting to note that one does not necessarily need to know the parameters $\alpha$ or $\beta$ precisely to implement this estimator.  In particular, because \eqref{eqn:sourceLocEstimator} is self-normalizing, the estimator automatically adapts to the source signal strength, $\alpha$. In addition, \cite{rabbat05} shows that this estimator is robust to choice of $\gamma$.  In particular, even if $\beta$ is not known precisely, the performance of \eqref{eqn:sourceLocEstimator} degrades gracefully.  On the other hand, the maximum likelihood approach is very sensitive to model mismatch and estimating $\alpha$ and $\beta$ can be challenging.

Note that \eqref{eqn:sourceLocEstimator} is a ratio of linear functions of the measurements at each node.  To compute \eqref{eqn:sourceLocEstimator}, we run two parallel instances of gossip over the network, one each for the numerator and the denominator.  If each node initializes $x^N_i(0) = y_i K(f_i)$, and $x^D_i(0) = K(f_i)$, then executing gossip iterations will cause the values at each node to converge to $\lim_{t \rightarrow \infty} x^N_i(t) = \frac{1}{n}\sum_{j=1}^n y_j K(f_j)$ and $\lim_{t \rightarrow \infty} x^D_i = \frac{1}{n} \sum_{j=1}^n K(f_j)$.  Of course, in a practical implementation one would stop gossiping after a fixed number of iterations, $t_{stop}$, which depends on the desired accuracy and network topology.  Then, each node can locally compute the estimate $x^N_i(t_{stop}) / x^D_i(t_{stop})$ of the source's location.  Note that throughout this section it was assumed that each node knows its own location.  This can also be accomplished using a gossip-style algorithm, as described in \cite{Khan09}.

\subsection{Distributed Compression and Field Estimation}
\label{subsec:fieldEst}

Extracting information in an energy-efficient and communication-efficient manner is a fundamental challenge in wireless sensor network systems.  In many cases, users are interested in gathering data to see an ``image" of activity or sensed values over the entire region.  Let $f_i \in \Re$ denote the measurement at node $i$, and let $f \in \Re^n$ denote the network signal obtained by stacking these values into a vector.  Having each sensor transmit $f_i$ directly to an information sink is inefficient in many situations.  In particular, when the values at different nodes are correlated or the signal is compressible, then one can transmit less data without loosing the salient information in the signal.  Distributed source coding approaches attempt to reduce the total number of bits transmitted by leveraging the celebrated results of Slepian and Wolf~\cite{slepian73} to code with side information \cite{servetto02,pradhan02}.  These approaches make assumptions about statistical characteristics of the underlying data distribution that may be difficult to verify in practice.

An alternative approach is based on linear transform coding, gossip algorithms, and compressive sensing.  It has been observed that many natural signals are compressible under some linear transformation.  That is, although $f$ may have energy in all locations (i.e., $f_i > 0$ for all $i$), there is a linear basis transformation matrix, $T\in \Re^{n\times n}$, such that when $f$ is represented in terms of the basis $T$ by computing $\theta = T f$, the transformed signal $\theta$ is compressible (i.e., $\theta_j \approx 0$ for many $j$).  For example, it is well known that smooth one-dimensional signals are well approximated using the Fourier basis, and piece-wise smooth images with smooth boundaries (a reasonable model for images) are well-approximated using wavelet bases \cite{mallat}.

\begin{figure*}
\centering %
\subfigure[][]{\includegraphics[width=2.4in]{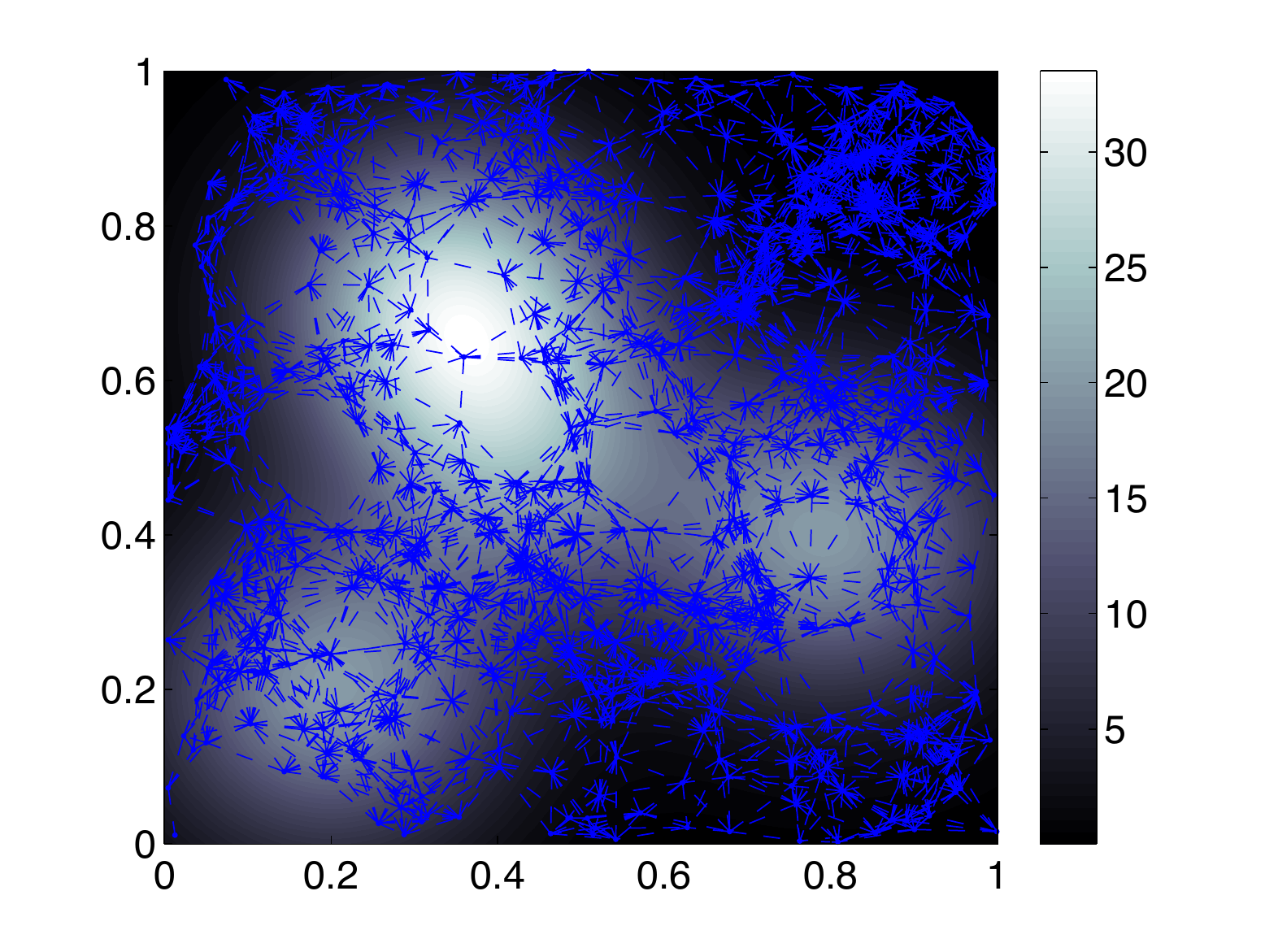}}%
\subfigure[][]{\includegraphics[width=2.4in]{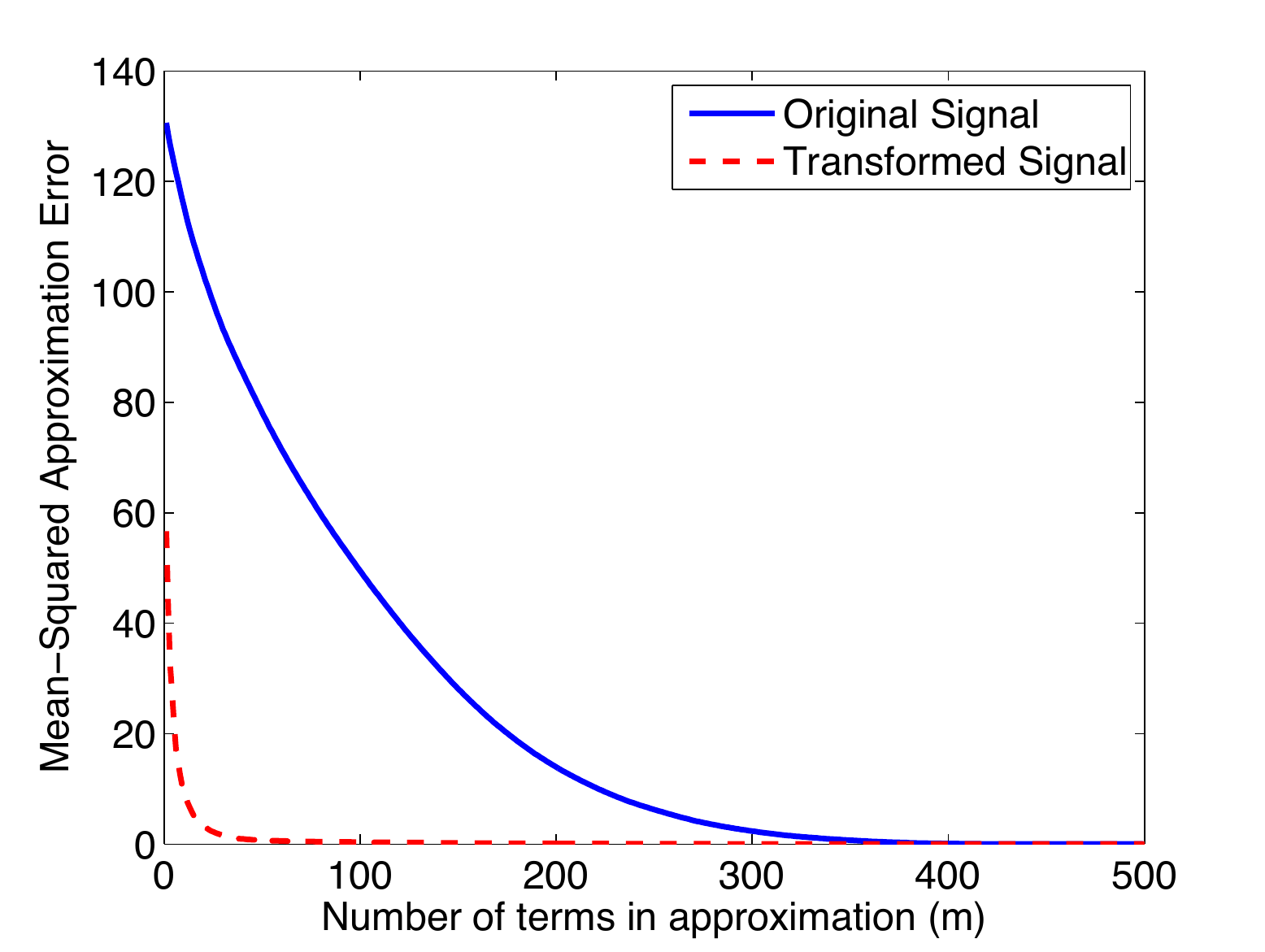}}%
\subfigure[][]{\includegraphics[width=2.4in]{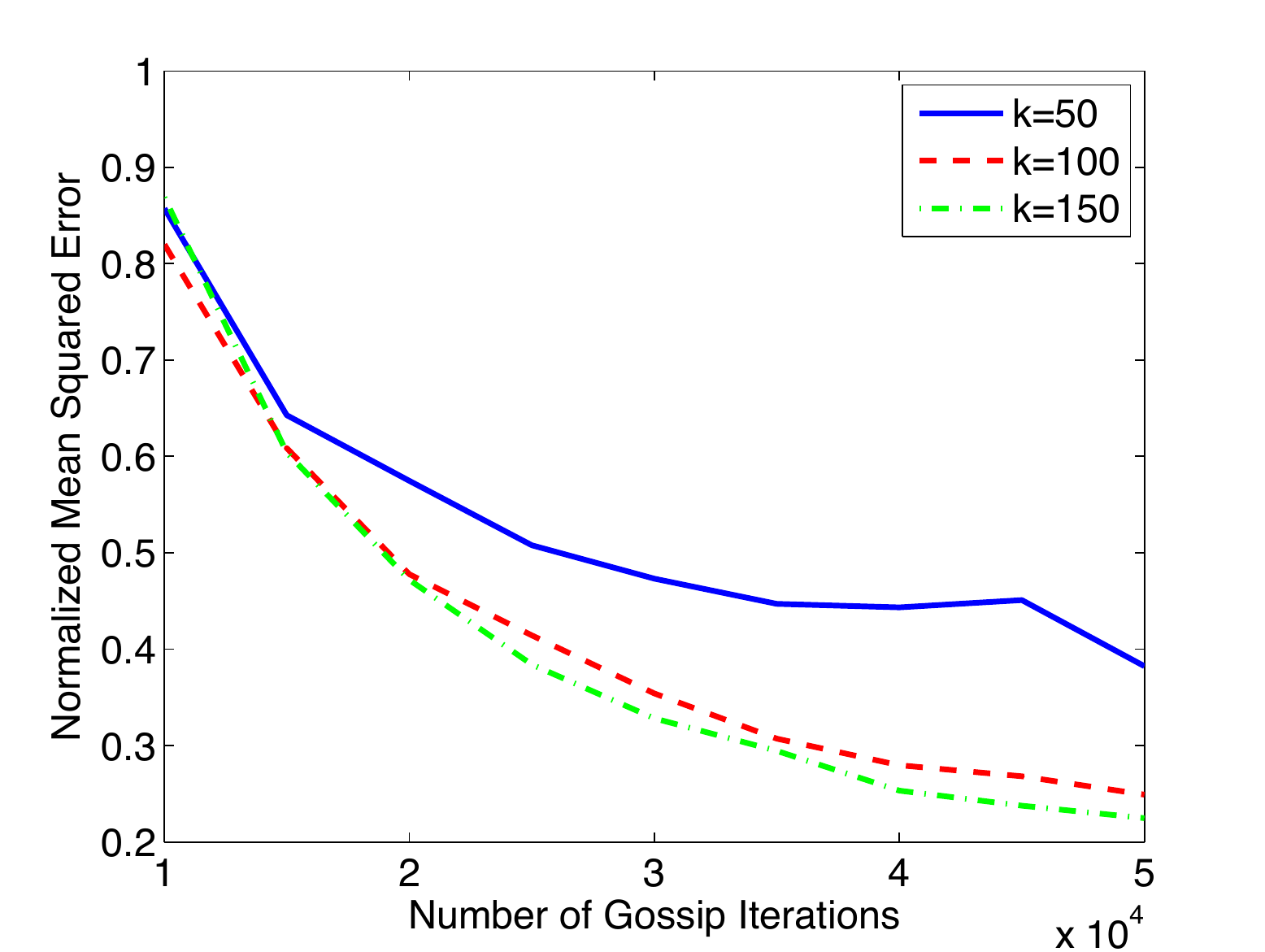}}%
\caption{Example illustrating compression of a smooth signal.  Panel (a) shows the original smooth signal which is sampled at 500 random node locations, and nodes are connected as in a random geometric graph.  Panel (b) illustrates the $m$-term approximation error decay in both the original basis and using the eigenvectors of the graph Laplacian as a transform, which is analogous to taking a Fourier transform of signals supported on the network.  Panel (c) illustrates the reconstruction error after gossiping on random linear combinations of the sensor measurements and reconstructing using compressed sensing techniques.  Note that using more random linear projections (larger $k$), gives lower error, but the number of projections used is much smaller than the network size.}
\label{fig:compressibleExample}
\end{figure*}

To formally capture the notion of compressibility using ideas from the theory of nonlinear approximation \cite{devore98}, we reorder the coefficients $\theta_j$ in order of decreasing magnitude,
\begin{equation}
|\theta_{(1)}| \ge |\theta_{(2)}| \ge |\theta_{(3)}| \ge \dots \ge |\theta_{(n)}|,
\end{equation}
and then define the best $m$-term approximation of $f$ in $T$ as $f^{(m)} = \sum_{j=1}^m \theta_{(j)} T_{:, (j)}$, where $T_{:,j}$ denotes the $j$th column of $T$.  This is analogous to projecting $f$ onto the $m$-dimensional subspace of $T$ that captures the most energy in $f$.  We then say that $f$ is $\alpha$-compressible in $T$, for $\alpha \ge 1$, when the mean squared approximation error behaves like
\begin{equation}
\frac{1}{n} \|f - f^{(m)}\|^2 \le C m^{-2\alpha},
\end{equation}
for some constant $C > 0$.  Since the error exhibits a power-law decay in $m$ for compressible signals, it is possible to achieve a small mean squared approximation error while only computing and/or communicating the few most significant coefficients $\theta_{(1)}, \dots, \theta_{(m)}$.  Figure~\ref{fig:compressibleExample} shows an example where 500 nodes forming a random geometric graph sample a smooth function.  As a compressing basis $T$, we use the eigenvectors of the normalized graph Laplacian (a function of the network topology), which are analogous to the Fourier basis vectors for signals supported on $G$ \cite{chung}.

Observe that each coefficient, $\theta_{j}$, is a linear function of the data at each node, and so one could conceivably compute these coefficients using gossip algorithms.  Assuming that each node $i$ knows the values $\{T_{j,i}\}_{j=1}^n$ in each basis vector, to compute $\theta_j$, we can initialize $x_i(0) = n T_{j,i} f_i$, and, by gossiping, each node will compute $\lim_{t \rightarrow \infty} x_i(t) = \sum_{k=1}^n T_{j,k} f_k = \theta_j$.  The main challenge with this approach is that the indices of the most significant coefficients are very signal-specific, and are generally not known in advance.

We can avoid this issue by making use of the recent theory of compressive sensing~\cite{candes:it05,donoho:it06,candes:it06b}, which says that one can recover sparse signals from a small collection of random linear combinations of the measurements.  In the present setting, to implement the gathering of $k$ compressive sensing measurements using gossip algorithms, each node initializes $k$ parallel instances of gossip with $x_{i,j}(0) = n A_{i,j} f_i$, $j=1,\dots,k$, where $A_{i,j}$, e.g., are i.i.d.~zero-mean normal random variables with variance $1/n$.  Let $\bar{x}_{j}$ denote the limiting value of the $j$th gossip instance at each node.  Stacking these into the vector, $\bar{x}$, any node can recover an estimate of the signal $f$ by solving the optimization
\begin{equation}
\min_{\theta} \|\bar{x} - A T^{T}\theta\|^2 + \tau \sum_{i=1}^n |\theta_i|,
\end{equation}
where $\tau > 0$ is a regularization parameter.  In practice, the values $A_{i,j}$ can be pseudo-randomly generated at each node using a predefined seeding mechanism.  Then, any user can retrieve the gossip values $\{x_{i,j}(t)\}_{j=1}^k$ from any node $i$ and solve the reconstruction.  Moreover, note that the compressing transformation $T$ only needs to be known at reconstruction time, and to initialize the gossip instances each node only needs its measurement and pseudo-randomly generated values $A_{i,j}$.  In general, there is a tradeoff between 1) $k$, the number of compressed sensing measurements collected, 2) $\epsilon$, the accuracy to which the gossip algorithm is run, 3) the number of transmissions required for this computation, and 4) the average reconstruction accuracy available at each node.  For an $\alpha$-compressible signal $f$, compressed sensing theory provides bounds on the mean squared reconstruction error as a function of $k$ and $\alpha$, assuming the values, $\bar{x}$, are calculated precisely.  Larger $k$ corresponds to lower error, and the error decays rapidly with $k$ (similar to the $m$-term approximation), so one can obtain a very accurate estimate of $f$ with $k \ll n$ measurements.  Inaccurate computation of the compressed sensing values, $\bar{x}$, due to gossiping for a finite number of iterations, can be thought of as adding noise to the values $\bar{x}$, and increases the overall reconstruction error.  Figure~\ref{fig:compressibleExample}(c) illustrates, via numerical simulation, the tradeoff between varying $k$ and the number of gossip iterations.  For more on the theoretical performance guarantees achievable in this formulation, see~\cite{rabbat06,haupt08}.  A gossip-based approach to solving the reconstruction problem in a distributed fashion is described in~\cite{Schmidt09}.  For an alternative approach to using gossip for distributed field estimation, see~\cite{Sarkar07}.

\section{Conclusion and Future Directions}
\label{sec:conc}

Because of their simplicity and robustness, gossip algorithms are an  
attractive approach to distributed in-network processing in wireless  
sensor networks, and this article surveyed recent results in this  
area.  A major concern in sensor networks revolves around conserving  
limited bandwidth and energy resources, and in the context of  
iterative gossip algorithms, this is directly related to the rate of  
convergence.  One thread of the discussion covered fast gossiping in  
wireless network topologies.  Another thread focused on understanding  
and designing for the effects of wireless transmission, including  
source and channel coding.  Finally, we have illustrated how gossip  
algorithms can be used for a diverse range of tasks, including  
estimation and compression.

Currently, this research is branching into a number of directions.   
One area of active research is investigating gossip algorithms that go  
beyond computing linear functions and averages.  Just as the average  
can be viewed as the minimizer of a quadratic cost function,  
researchers are studying what other classes of functions can be  
optimized within the gossip framework~\cite{Ram09}.  A related direction is  
investigating the connections between gossip algorithms and message-passing algorithms for distributed inference and information fusion,  
such as belief propagation~\cite{cetin06,Moallemi06}.
While it is clear that computing pairwise averages is similar to the  
sum-product algorithm for computing marginals of distributions, there  
is no explicit connection between these families of distributed  
algorithms. It would be interesting to demonstrate that pairwise  
gossip and its generalizations correspond to messages of the sum- 
product (or max-product) algorithm for an appropriate Markov random  
field. Such potentials would guarantee convergence (which is not  
guaranteed in general iterative message-passing) and further establish  
explicit convergence and message scheduling results.

Another interesting research direction involves understanding the  
effects of intermittent links and dynamic topologies, and in  
particular the effects of node mobility. Early work~\cite{Sarwate08}  
has analyzed i.i.d mobility models and shown that
mobility can greatly benefit convergence under some conditions.  
Generalizing to more realistic
mobility models seems to be a very interesting research direction that  
would also be relevant in practice since gossip algorithms are more  
useful in such dynamic environments.

Gossip algorithms are certainly relevant in other applications that  
arise in social networks and the interaction of mobile devices with  
social networks. Distributed inference and information fusion in such  
dynamic networked environments is certainly going to pose substantial  
challenges for future research.

\bibliographystyle{IEEEtran}
\bibliography{gossipMike,gossipAnna,gossipAlex,gossipSoummyaJose}

\end{document}